\pdfoutput=1
\RequirePackage{fix-cm}
\documentclass{article}
\usepackage{arxiv}
\usepackage{graphicx}
\usepackage{subcaption}
\usepackage{url}
\usepackage{bm}
\usepackage{amsmath}
\usepackage[inline]{asymptote}
\usepackage{import}
\usepackage{siunitx}
\usepackage{hyperref}
\usepackage{algorithm}
\usepackage{algpseudocode}

\usepackage{xcolor}

\begin{document}
\title{Quantitative Evaluation of SPH in TIG Spot Welding
}


\author{Stefan Rhys Jeske\\
        RWTH Aachen University\\
        \texttt{jeske@cs.rwth-aachen.de}\\ 
        \And
        Marek Simon\\
        RWTH Aachen University\\
        \texttt{simon@isf.rwth-aachen.de}\\     
        \And
        Oleksii Semenov\\
        Paton Welding Insitute\\
        \texttt{oleksii.semenov@ukr.net}\\     
        \And
        Jan Kruska\\
        RWTH Aachen University\\
        \texttt{jan.kruska@isf.rwth-aachen.de}\\    
        \And
        Oleg Mokrov\\
        RWTH Aachen University\\
        \texttt{oleg.mokrov@isf.rwth-aachen.de}\\    
        \And
        Rahul Sharma\\
        RWTH Aachen University\\
        \texttt{sharma@isf.rwth-aachen.de}\\    
        \And
        Uwe Reisgen\\
        RWTH Aachen University\\
        \texttt{reisgen@isf.rwth-aachen.de}\\    
        \And
        Jan Bender\\
        RWTH Aachen University\\
        \texttt{bender@cs.rwth-aachen.de}\\    
}



\maketitle

\begin{abstract}
While the application of the Smoothed Particle Hydrodynamics (SPH) method for the modeling of welding processes has become increasingly popular in recent years, little is yet known about the quantitative predictive capability of this method.
We propose a novel SPH model for the simulation of the tungsten inert gas (TIG) spot welding process and conduct a thorough comparison between our SPH implementation and two Finite Element Method (FEM) based models.
In order to be able to quantitatively compare the results of our SPH simulation method with grid based methods we additionally propose an improved particle to grid interpolation method based on linear least-squares with an optional hole-filling pass which accounts for missing particles.
We show that SPH is able to yield excellent results, especially given the observed deviations between the investigated FEM methods and as such, we validate the accuracy of the method for an industrially relevant engineering application.

\end{abstract}

\section{Introduction}
\label{sec:intro}
The Smoothed Particle Hydrodynamics method (SPH) is a Lagrangian discretization method, which has been investigated in recent years for the simulation of manufacturing processes, such as arc welding processes, see  \cite{THF18} \cite{INI15} \cite{IIFS11} \cite{KTT20}.
However, until now, the method has not been quantitatively compared for the application to arc welding processes, to asses its performance regarding calculation speed and computational accuracy. 
Therefore, in this work, the open source implementation of the SPH method SPlisHSPlasH \cite{splishsplash} has been extended to allow for the consideration of heat transfer and respective boundary conditions as well as the influence of external forces (such as the Lorentz force).
In order to study the applicability of our SPH method for the simulation of arc welding processes, a test case was set up modeling a simplified spot tungsten inert gas (TIG) welding process. The test problem was solved separately by means of our SPH implementation and two different FEM models implemented in Wolfram Mathematica (WM) \cite{semenov2021} and COMSOL Multiphysics, allowing for the simulation of weld pool formation under the axisymmetric assumption.

For this comparison, the problem was idealized by using constant material properties in the simulation and assuming a flat free surface.
Typically, this kind of problem is well within the strengths of Eulerian methods, whereas the SPH method is known to have difficulties with strictly bounded domains.
While other works only consider the application of SPH to the entire process, we aim to first establish the accuracy of the SPH method in this idealized and bounded scenario. 
This is done by rigorously comparing against known Eulerian methods, whose properties have been well studied for many years.
The electromagnetic Lorentz force, which is the main driving force for the convective heat transfer in the melt pool, was calculated in advance by the WM-Code. 
The resulting vector field was exported and applied to the SPH simulation as an axisymmetric external force, while it was solved within COMSOL given the same boundary conditions.

We summarize our main contributions as follows:
\begin{itemize}
    \item The proposition of a novel SPH algorithm for the simulation of the TIG spot-welding process, which utilizes implicit time integration to significantly improve simulation stability, and enables the use of larger time steps.
    \item The construction and implementation of a fast projection scheme for the integration of surface effects for SPH fluids.
    \item A novel particle to grid interpolation method which, compared to existing methods such as Shepard interpolation, is able to generate smoother fields while maintaining very high interpolation accuracy. This is extended by a smoothing algorithm in order to fill in holes of the interpolated grid due to missing particles.
    \item The quantitative comparison and in depth analysis of our SPH method with the WM-FEM and the COMSOL-FEM on the simulation of TIG spot-welding.
\end{itemize}

Overall we are able to show remarkable agreement between our SPH method and the two FEM methods, especially the solution of COMSOL, while at the same time being entirely in the margin of error exhibited by the comparison of the WM-FEM method and the COMSOL-FEM method.
This result is made all the more remarkable considering that our SPH model solves the problem in full 3D while both WM and COMSOL solve the problem in 2D with rotational symmetry.

\subsection{Modeling of TIG Spot-Welding}

The tungsten inert-gas (TIG) welding processes is a widely applied welding process, that, while it has a lower productivity than gas-metal-arc welding processes, has its advantages when it comes to high requirements on weld seam quality and optical appearance. It is especially suitable for welding root passes, thin metal sheets and materials prone to oxidation. It is therefore often applied in chemical or food processing plants.

For the present validation the ``spot-welding'' variant of the process is being studied for simplicity reasons, where the welding torch position is fixed and no filler material is supplied.
In addition, we consider high frequency TIG (HF-TIG), in which the electromagnetic Lorentz force, as a result of the concentration of the arc, causes the formation of significant flow patterns in the weld pool.
For the usual process, in which the distribution of electric current on the surface of the liquid is typically less concentrated, the Lorentz force is less dominant. 
Therefore, only the weld pool formation caused by the heat flux from the arc and the influence of the Lorentz force is studied, which is sufficient for a first qualitative comparison with Eulerian methods.

To facilitate the comparison, the non-linearities of the material parameters, like latent heat and temperature dependent thermal and electrical conductivity or viscosity were neglected, instead applying constant values for the relevant material parameters, which allowed the Lorentz force to be pre-computed.
Also, since the modeling of the free surface of the weld pool requires a special treatment in the Eulerian framework, like volume of fluid (VOF) or level-set, the free surface deformation and Marangoni effect were neglected, although their modeling would pose a great advantage for the SPH method.
The main focus of the present work is therefore not to present an accurate modeling of the welding process, but rather to investigate the applicability of the SPH method for the conditions present in this process and to quantitatively compare its performance against the common Eulerian methods, under the same conditions. 

\subsection{SPH Methods in Welding Simulation}

The Smoothed Particle Hydrodynamics (SPH) method was originally proposed by Lucy \cite{Luc77} and Gingold and Monaghan \cite{GM77} in the field of Astrophysics.
Since then it has been adopted for many different applications, including the simulation of weld pool dynamics.
The mesh-free nature of SPH enables the simulation of large deformations and free surface motion and coupling of many physical processes, which makes it an attractive method for many real world problems. 

Das and Cleary \cite{DC16} use SPH in three-dimensional arc welding simulations in order to study temperature distributions, flow patterns and plastic strain in the filler material and residual thermal stresses in the work piece.

Ito et al.~\cite{INI15} perform full simulations of TIG welding using the SPH method and show results for different material properties due to different sulfur contents and evaluate the flow patterns and the shape of the weld pool.

Trautmann et al.~\cite{THF17} similarly perform weld pool simulations using SPH for a TIG welding process.
They consider buoyancy, viscosity and surface tension as flow driving forces and the arc pressure, shear and thermal effect were parametrized using experimental studies.
The penetration profiles of three different welding currents were compared to experimental results and show decent agreement.  

A hybrid approach is investigated by Komen et al. \cite{KTT20}, who use a Eulerian grid and Lagrangian particles in conjunction in order to simulate gas metal arc welding (GMAW).
Molten metal is simulated by means of SPH, while the arc plasma and gas are simulated on a grid.
The methods are then weakly coupled and executed iteratively.

Komen et al.~\cite{KST18} also simulate the GMAW process under consideration of droplet formation and compare the weld pool shapes against experimental results.
In order to more easily visualize results they use an ensemble averaging in order to transfer particle data onto a regular grid.

These types of simulations are difficult to validate, as they describe very complex systems with many interacting components, and this is evident with Trautmann et al.~\cite{THF17} being one of the few works which attempts to validate their results using experimental data.
In this paper we chose instead to compare our proposed SPH method against Eulerian simulations, as a proof of concept which shows that SPH is able to obtain very good agreement in the resulting weld pool shapes, as well as temperature and velocity distributions.
This allows us to justify the usage of SPH to obtain physically meaningful results, also for cases that pose more difficulty for Eulerian methods, for example when considering a free surface. 
The performed investigations confirm the quantitative accuracy of our proposed method, as well as the SPH method as a whole for these kinds of applications.

\subsection{Eulerian Methods in Welding Simulation}

Eulerian methods are very common for the simulation of arc welding processes and have been used since the advent of computational welding simulation. 
A common difficulty in these methods is the calculation of free surface flows. 
There are several numerical approaches which are commonly used for description of these free surface flows. 
They can be divided into two main groups \cite{semenov2014}. 
The first group includes the so called Front Capturing Methods (FCM) in which a fixed Eulerian computational mesh is employed and a free surface is "expanded" along the volume of a certain layer. 
The thickness of this layer corresponds to several lengths of a computational cell. 
The most popular FCM are Volume of Fluid (VOF) \cite{haidar1996},\cite{hertel2014} and Level Set (LS) \cite{CCC+20} methods. 

There exist also a number of other approaches where the free surface is considered as a sharp interface between two media, e.g. \cite{nguyen2017}, \cite{medale2004}, \cite{medale2008}. There the Ar\-bitrary-Lagrangeian-Eulerian method (ALE) is used, which allows for a deformation of the mesh, but it does not allow to solve problems with significant topological changes, like flow-splitting. However, most common arc welding processes involve a melting, detachment and an impingement of a filler material into a weld pool.  
Additionally, the attachment of the electric arc to the liquid electrodes is determined by thin sheaths (anode and cathode sheaths), which are located exactly at the surface of the liquid electrodes and therefore a sharp definition of the free surface is necessary to consider these phenomena \cite{MSS+21}, \cite{SDK+12}. 
Therefore, neither the VOF/LS approach, nor the ALE approach are completely satisfactory to accurately capture the process. Due to its strengths in modeling free surfaces as well as considering discontinuities and large topological changes and deformations, the SPH method became an interesting approach for modeling arc welding processes. However, little is yet known about the quantitative performance for the calculation of conductive/convective heat transfer compared to the established Eulerian methods, in the context of arc welding processes.


\section{Method}
\label{sec:sph}
In this section we briefly describe our SPH model and discretization of the
governing equations for incompressible weld pool dynamics.
The SPH method discretizes mass at particles in space.
These particles are advected and tracked through time and carry associated field quantities with them.
The SPH method uses interpolation in order to compute unknown quantities and deri\-vatives needed to solve Partial Differential Equations (PDEs).
An arbitrary scalar quantity $A(\bm{x})$ can thus be computed by interpolation from surrounding particles
\begin{equation}
\label{eq:sph_sum}
A(\bm{x}) = \sum_{j\in \mathcal{N}_x} V_j A_j W(\bm{x} - \bm{x}_j; h),
\end{equation}
where $W(\bm{x} - \bm{x}_j; h)$ is a compactly supported weighting function around position $\bm{x} \in \mathcal{R}^3$ with smoothing length $h$. 
The term $\sum_{j\in \mathcal{N}_x}$ denotes a summation over the neighboring particles $j$ of position $\bm{x}$, which lie within the compact support radius of weighting function $W(\bm{x})$.
Derivative operators are typically shifted to the kernel function $W$ which significantly simplifies the discretization of PDEs.
For more information about the background of the SPH method the reader is referred to Koschier et al.~\cite{KBST19} and Price \cite{Pri12a}.

\subsection{Fluid Flow}
\label{sec:fluids}
The Navier-Stokes equations for incompressible fluid flow are given by
\begin{equation}
    \label{eq:navier_stokes}
    \rho \frac{D\bm{v}}{D t} = -\nabla p + \mu \nabla^2 \bm{v} + \bm{f}_{ext}.
\end{equation}
Here $\rho$ denotes the fluid density [kg m$^{-3}$], $\bm{v}$ the velocity [m s$^{-1}$], $p$ the pressure [N m$^{-2}$], $\mu$ the dynamic viscosity [Pa s] and $\bm{f}_{ext}$ the external volumetric forces [N m$^{-3}$]. 
In addition to the Navier-Stokes equations, the continuity equation has to be fulfilled
\begin{equation}
    \frac{D \rho}{D t} = 0,
\end{equation}
which states that there must be no change in density, a necessary condition for incompressible fluids.
This equation is also often used in convective form which relates the local density change to the divergence of the velocity field.
\begin{equation}
    \frac{\partial \rho}{\partial t} = -\rho \nabla \cdot \bm{v} = 0
\end{equation}
The equations are solved using operator splitting.
First the density of all particles $i,\, \forall i \in [1, N]$, where $N$ is the total number of particles in the simulation, is computed using the SPH interpolation function
\begin{equation}
    \rho_i =  \sum_{j\in \mathcal{N}_i} m_j W_{ij},
\end{equation}
where $W_{ij}$ is shorthand for $W(\bm{x}_i - \bm{x}_j; h)$. 
Afterwards the acceleration due to pressure forces is computed using the Divergence-Free SPH (DFSPH) method, as presented by Bender and Koschier \cite{BK15}.
Using this method, a set of momentum conserving forces is obtained which guarantees both constant density as well as a divergence-free velocity field.
Other established methods, such as advection of density using the continuity equation \cite{DC16} or explicit computation of pressure forces using a state equation \cite{BT07} may suffer from volume losses in the former case and typically require small simulation time steps for the latter in order to ensure incompressibility.
In contrast to this, we have found DFSPH, which implicitly computes pressure forces, to perform very well in all conditions and it enabled the usage of large time steps while still remaining stable.

The viscosity term is also computed implicitly using the viscosity model by Weiler et al.~\cite{WKBB18}. 
This method computes accelerations 
\begin{equation}
  \bm{a}_{i} = \frac{\bm{v}_i^{t+1} - \bm{v}_i^t}{\Delta t}
\end{equation}
due to viscous forces, by solving for $\bm{v}_\mu^{t+1}$ in the following equation
\begin{equation}
    \bm{v}^{t+1} = \bm{v}^t + \Delta t \frac{\mu}{\rho} \nabla^2 \bm{v}^{t+1}.
    \label{eq:viscosity}
\end{equation}

This can be written as a square symmetric positive definite linear system and solved efficiently in a matrix-free context using the conjugate gradient method.
To the best of our knowledge, implicit viscosity solvers have not yet been employed for the simulation of weld pool dynamics and we have found this model to also be very stable when using large time steps.
Further discussion will follow in Section \ref{sec:discussion}.

The remaining forces in Equation \eqref{eq:navier_stokes} which are part of the external volumetric force are the Lorentz force and the Momentum Sink which accounts for the morphological material change during solidification and mel\-ting.

The Lorentz force is computed in advance of the simulation and interpolated bilinearly from a grid onto the particles. This is analogous to the widely used approach for calculating Lorentz forces in arc welding process simulation, as for example used by Cho and Na \cite{CN21}, where also the stationary analytical solution developed by Kou and Sun \cite{KS85} is used to calculate the Lorentz forces.

The Momentum Sink is computed using the Darcy-term method, which is a common approach in literature to model the solidification of pure metals \cite{BVR88} and alloys \cite{VBP90}, which is also a common approach in arc welding simulation, see e.g. \cite{CN21}. The method adds an acceleration to the Navier-Stokes equation which has a strong flow inhibiting effect on the fluid, once the temperature of the fluid reaches below melting temperature $T_{l}$
\begin{align}
\bm{a}_{porosity} &= -\bm{v} C (1-f_l(T)) \\
f_l(T) &= \begin{cases}
1 & T > T_l\\
0 & T \leq T_l.
\end{cases}
\end{align}

In the above equations $C$ denotes the morphological constant and $f_l(T)$ is the temperature dependent liquid fraction, which is modeled using the heaviside function.
The values for $C$ are often very large, resulting in very large deceleration as soon as $T\leq T_l$, so large in fact that simulations using explicit time stepping can become unstable.
We remedy this by constructing an algebraic equation which computes the acceleration using the projected velocity of the next time step as in the following
\begin{align}
    \bm{a}_{porosity} = \frac{\bm{v}^{t+1} - \bm{v}^{t}}{\Delta t} &= -\bm{v}^{t+1} C (1-f_l(T))\\
    \bm{v}^{t+1} &= \bm{v}^t \frac{1}{1 + C \Delta t (1-f_l(T))}.
\end{align}
The acceleration is then computed by inserting the expression for $\bm{v}^{t+1}$
\begin{equation}
    \bm{a}_{porosity} = \frac{\bm{v}^t}{\Delta t}\left( \frac{1}{1 + C \Delta t  (1-f_l(T))} - 1\right).
    \label{eq:porosity}
\end{equation}
This semi-implicit formulation allows the usage of larger time steps in the simulation without causing instabilities.
It should be noted that using large time steps with implicit Euler integration is known to add a significant damping effect, which diminishes the effect of the Darcy-term.
This can be controlled by carefully determining a time step in the order of magnitude of $C^{-1}$ which is typically still very small, $\mathcal{O}$(\SIrange{e-8}{e-4}{\second}), but not necessarily as small as required for explicit integration.

The main fluid driving forces in our simplified model of TIG spot welding is the Lorentz force while the fluid behavior is otherwise significantly influenced by viscosity and the melting of solid material.
The extensive use of implicit solvers enables us to have stable simulations when using large simulation time steps as well as large morphological constants $C$.

Finally, an overview of our SPH fluid solver is presented in Algorithm \ref{lst:fluid_solver}.

\begin{algorithm}[t]
  \caption{Fluid Solver}\label{lst:fluid_solver}
  \begin{algorithmic}[1]
    \Procedure{FluidSolve}{}
    \State\Call{SolveDivergenceFreeVelocity}{$i$}\Comment{\cite{BK15}}
    \State\Call{SolveViscosity}{$i$}\Comment{Eq. \eqref{eq:viscosity}}
    \For{all particles $i$}
      \State\Call{SolveMomentumSink}{$i$}\Comment{Eq. \eqref{eq:porosity}}
    \EndFor
    \State\Call{SolveConstantDensityPressure}{$i$}\Comment{\cite{BK15}}
    \State $\Delta t \gets$ \Call{ComputeCFLTimeStep}{\,}
    \For{all particles $i$}
      \State $\bm{v}_i \gets \bm{v}_i + \Delta t \left(\bm{a}_i + \frac{\bm{f}_{ext}}{\rho}\right)$
      \State $\bm{x}_i \gets \bm{x}_i + \Delta t \bm{v}_i$
    \EndFor
    \EndProcedure
  \end{algorithmic}
\end{algorithm}

\subsection{Heat Transfer}
\label{sec:method_heat_transfer}

Heat transfer is computed using the Fourier equation
\begin{equation}
    \frac{D(\rho c_p T)}{D t} =  \nabla\cdot\left(\lambda \nabla T \right) + \dot{q}''',
    \label{eq:energy_transport}
\end{equation}
where $\rho$ denotes the material density {[}kg m$^{-3}${]}, $c_p$ the specific heat capacity {[}J kg$^{-1}$ K$^{-1}${]}, $T$ the temperature {[}K{]}, $\lambda$ the thermal conductivity {[}W K$^{-1}$ m$^{-1}${]} and $\dot{q}'''$ the contribution from volumetric heat sources {[}W m$^{-3}${]}. 
For the purpose of our SPH simulation we make use of the relationship between specific enthalpy $h$ {[}J kg$^{-1}${]} and the temperature when the observed medium is incompressible
\begin{equation}
    h(T) = \int_0^T c_p(T) dT.
\label{eq:enthalpy_temperature}
\end{equation}
In this formulation $c_p$ may also be a function of temperature, taking into account, e.g. the latent heat of melting. 
Resulting in the following formulation in terms of specific enthalpy
\begin{equation}
    h\frac{D \rho}{D t} + \rho\frac{D h}{D t} =  \nabla\cdot\left(\lambda \nabla T \right) + \dot{q}'''.
\end{equation}
$\frac{D\rho}{Dt} = 0$ is already enforced by the constant density component of the implicit pressure solver so that this term immediately drops out. 
The above equation is discretized using SPH and explicit Euler time integration, resulting in the following discrete equation for the fluid particle with index $i$
\begin{equation}
    \rho_i \frac{h_i^{t+1} - h_i^t}{\Delta t} = \nabla \cdot (\lambda \nabla T)_i^t + \dot{q}_i^{t}{}'''.
    \label{eq:heat_conduction_integration}
\end{equation}
The heat conduction term is discretized by
\begin{equation}
    \nabla \cdot (\lambda \nabla T)_i = \sum_{j\in \mathcal{N}_i}\frac{m_j}{ \rho_j}\frac{4\lambda_i\lambda_j}{\lambda_i + \lambda_j}(T_i - T_j)\frac{ \nabla_i W_{ij}\cdot \bm{r}_{ij}}{||\bm{r}_{ij}||^2}
    \label{eq:heat_conduction}
\end{equation}
as proposed by Brookshaw \cite{Bro85} and is well documented in the literature, e.g. by Trautmann et al.~\cite{THF17} and Das and Cleary \cite{DC16}.
It should be noted that the thermal conductivity $\lambda_i = \lambda(T_i)$ is generally a function of temperature and $\bm{r}_{ij} = \bm{x}_i - \bm{x}_j$.

\subsection{Heat Sources}
\label{sec:heat_sources}

\begin{figure}
    \centering
    \def\svgwidth{0.45\textwidth} 
\begingroup%
  \makeatletter%
  \providecommand\color[2][]{%
    \errmessage{(Inkscape) Color is used for the text in Inkscape, but the package 'color.sty' is not loaded}%
    \renewcommand\color[2][]{}%
  }%
  \providecommand\transparent[1]{%
    \errmessage{(Inkscape) Transparency is used (non-zero) for the text in Inkscape, but the package 'transparent.sty' is not loaded}%
    \renewcommand\transparent[1]{}%
  }%
  \providecommand\rotatebox[2]{#2}%
  \newcommand*\fsize{\dimexpr\f@size pt\relax}%
  \newcommand*\lineheight[1]{\fontsize{\fsize}{#1\fsize}\selectfont}%
  \ifx\svgwidth\undefined%
    \setlength{\unitlength}{445.98514587bp}%
    \ifx\svgscale\undefined%
      \relax%
    \else%
      \setlength{\unitlength}{\unitlength * \real{\svgscale}}%
    \fi%
  \else%
    \setlength{\unitlength}{\svgwidth}%
  \fi%
  \global\let\svgwidth\undefined%
  \global\let\svgscale\undefined%
  \makeatother%
  \begin{picture}(1,0.47754131)%
    \lineheight{1}%
    \setlength\tabcolsep{0pt}%
    \put(0,0){\includegraphics[width=\unitlength]{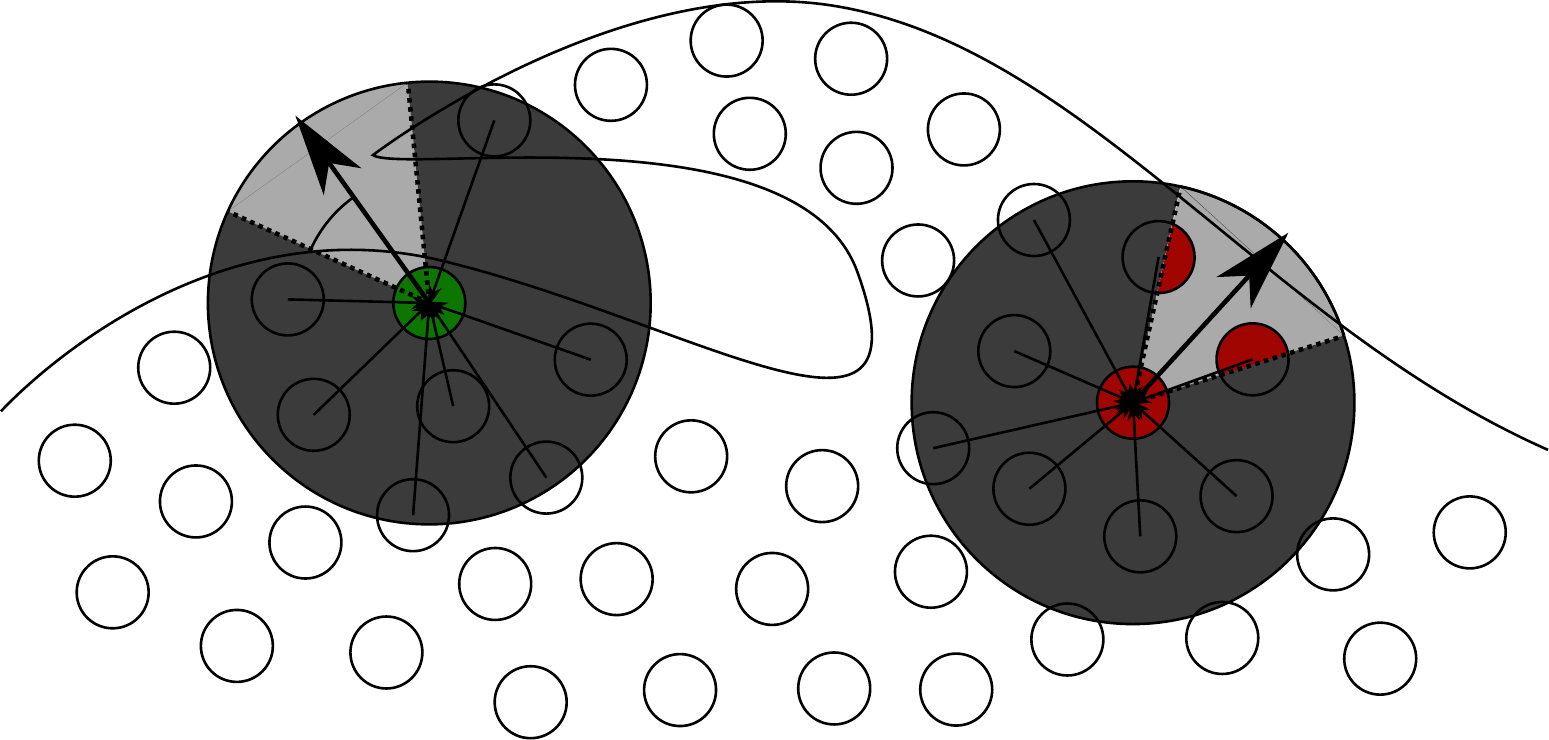}}%
    \put(0.17947073,0.4195742){\color[rgb]{0,0,0}\makebox(0,0)[lt]{\lineheight{1.25}\smash{\begin{tabular}[t]{l}$\hat{\mathbf{n}}_i$\end{tabular}}}}%
    \put(0.19009184,0.33426702){\color[rgb]{0,0,0}\makebox(0,0)[t]{\lineheight{1.25}\smash{\begin{tabular}[t]{c}$\varphi$\end{tabular}}}}%
    \put(0.30522325,0.33209777){\makebox(0,0)[lt]{\lineheight{1.25}\smash{\begin{tabular}[t]{l}$\mathbf{x}_i - \mathbf{x}_j$\end{tabular}}}}%
    \put(0.85312691,0.33737343){\color[rgb]{0,0,0}\makebox(0,0)[lt]{\lineheight{1.25}\smash{\begin{tabular}[t]{l}$\hat{\mathbf{n}}_i$\end{tabular}}}}%
    \put(0.76863041,0.19389485){\makebox(0,0)[lt]{\lineheight{1.25}\smash{\begin{tabular}[t]{l}$\mathbf{x}_i - \mathbf{x}_j$\end{tabular}}}}%
  \end{picture}%
\endgroup%

    \caption{Surface particle classification using Eq. \eqref{eq:surface_classification}. The green particle is classified as a surface particle while the red particle is classified as an interior particle.}
    \label{fig:method_coverage_vector_surface}
\end{figure}
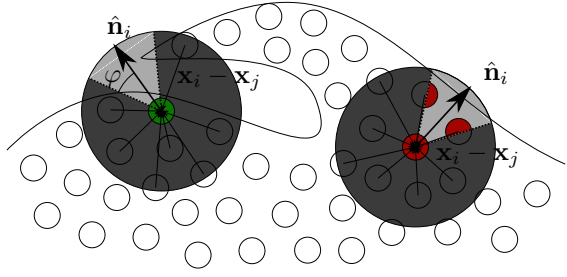

In this work, a surface heat source $\dot{q}_i^{t}{}'''$ with constant power and Gaussian distribution was used. 
There are many available surface classification techniques available, however the more accurate they are, the more computationally expensive they typically become. 
In this work it was found, that the coverage vector technique, as presented by Barecasco et al.~\cite{BTN13}, works quite well, since this results in almost flawless classification for the initial grid configuration and remains stable throughout. 
The normalized coverage vector $\hat{\bm{n}}_i$ for particle $i$ is computed as
\begin{align}
\bm{n}_i &= \sum_{j\in \mathcal{N}_i} (\bm{x}_i - \bm{x}_j) \\
\hat{\bm{n}}_i &= \frac{\bm{n}_i}{||\bm{n}_i||}.
\end{align}
This vector points in the direction of the lowest particle density in the particle neighborhood. 
Particle $i$ is classified as a surface particle, if there is no other particle $j$ in the direction of $\hat{\bm{n}}_i$ in a cone with an angle of
\begin{equation}
    \varphi < \cos^{-1}\left(\bm{\hat{n}_i} \cdot \frac{(\bm{x}_j - \bm{x}_i)}{||\bm{x}_j - \bm{x}_i||}\right), \forall j \in \mathcal{N}_i.
    \label{eq:surface_classification}
\end{equation}
The angle $\varphi$ is set to a fixed value. 
We have observed excellent classification results using an angle of $\varphi$ = \SI{35}{\degree}.
Here $\mathcal{N}_i$ contains the indices of particles in the compact support radius of particle $i$. 
Surface classification using the coverage vector is displayed in Figure \ref{fig:method_coverage_vector_surface}.
The particle shaded in green is classified as a surface particle, because no other particles are present in a cone of $\varphi$ around the coverage vector $\bm{\hat{n}_i}$.
The red particle is classified as an interior particle since there are two other particles in the cone around the coverage vector.

\begin{figure}
    \centering
    \includegraphics[width=.45\textwidth]{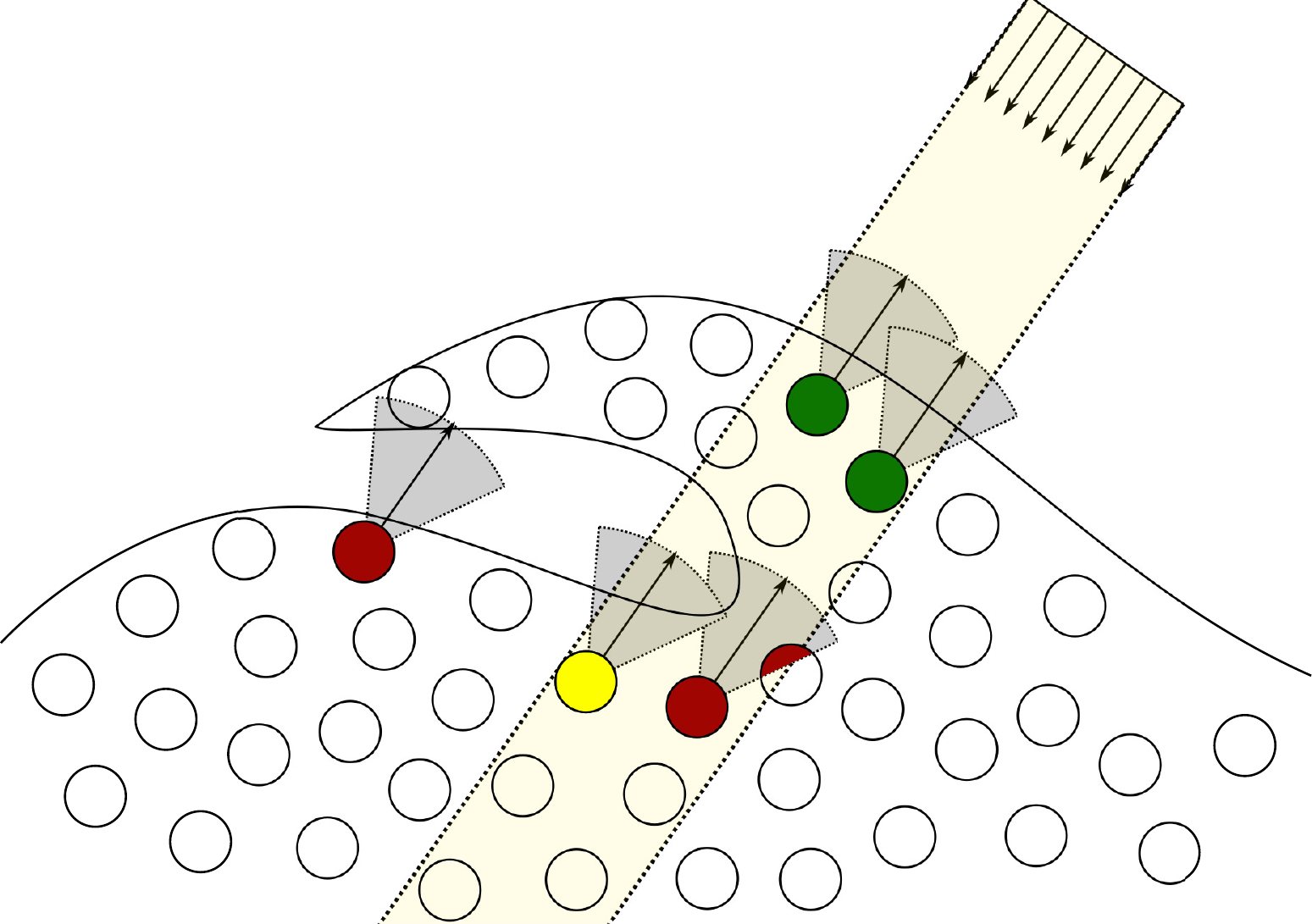}
    \caption{Projection of an area heat source onto previously detected surface particles. The green particles are visible, the yellow particle is discarded by a depth test, and the red particles are invisible to the heat source.}
    \label{fig:method_heat_source_projection}
\end{figure}

For the particles classified as surface particles, we propose the following fast projection method in order to compute the influence of a directed heat source.
In general however, this method could also be used to apply arbitrary boundary conditions at the fluid surface.
The heat source is computed by additionally computing the visibility of the surface particles from an area heat source.
This is shown in Figure \ref{fig:method_heat_source_projection}.
All surface particles outside of the region which lies in normal direction of the area source (white area) are discarded.
Then all remaining surface particles are checked for other neighboring particles in a cone around the vector in normal direction to the source.
The green particles are eventually classified as being visible from the surface while the red one is covered by a neighboring particle in the direction of the source.
The yellow particle would be classified as visible, but can be discarded by an additional depth test.

In this work, the heat source is parametrized as a Gaussian and in order to ensure constant power of the heat source, weighted normalization is applied using the previously computed Gaussian weights. 
This results in the expression
\begin{equation}
\dot{q}_{i,surface}'' = P_{source} w_i \frac{3}{\pi r_{HS}^2}\exp\left(-3\frac{||\bm{x_i} - \bm{x_0}||^2}{r_{HS}^2}\right),
\label{eq:heat_source}
\end{equation}
where $r_{HS}$ is the standard deviation, $\bm{x_0}$ the center of the Gaussian and $w_i$ the weight which is needed for normalization of the total power to the target target power $P_{source}$.

The subscript $i,surface$ denotes only the particles which have in the previous step been identified as surface particles and have been determined to be ``visible'' from the heat source, while $\dot{q}''$ denotes the heat source as Watts per unit area [\si{\W\per\m\squared}]. 

In our TIG simulation setup, all boundaries are adiabatic and do not require any special handling for heat transfer.
Since heat can only be conducted within the material itself, the SPH formulation is adiabatic by construction. 
Finally, the enthalpy in terms of the temperature (and vice versa) is precomputed by integration of Equation \eqref{eq:enthalpy_temperature}.
Our heat solver algorithm is outlined in Algorithm \ref{lst:compute_heat_transfer}, while the full algorithm containing both the heat solver and fluid solver is shown in Algorithm \ref{lst:sph_solver}.

\begin{algorithm}[t]
  \caption{Heat Transfer Computation}\label{lst:compute_heat_transfer}
  \begin{algorithmic}[1]
    \Procedure{ComputeHeatTransfer}{}
    \For{all particles $i$}
      \State\Call{ComputeTemperature}{$i$}\Comment{Eq. \eqref{eq:enthalpy_temperature}}
    \EndFor
    \For{all particles $i$}
      \State\Call{ClassifySurface}{$i$}\Comment{Eq. \eqref{eq:surface_classification}}
    \EndFor
    \For{all particles $i$}
      \State\Call{ApplyHeatSource}{$i$}\Comment{Eq. \eqref{eq:heat_source}}
    \EndFor
    \For{all particles $i$}
      \State\Call{ComputeHeatConduction}{$i$}\Comment{Eq. \eqref{eq:heat_conduction}}
    \EndFor
    \For{all particles $i$}
      \State\Call{ExplicitEulerIntegration}{$i$}\Comment{Eq. \eqref{eq:heat_conduction_integration}}
    \EndFor
    \EndProcedure
  \end{algorithmic}
\end{algorithm}

\begin{algorithm}[t]
  \caption{SPH Solver}\label{lst:sph_solver}
  \begin{algorithmic}[1]
    \Procedure{SolveFluidAndHeatConduction}{}
    \State $t \gets t_\text{start}$
    \While{$t < t_\text{end}$}
      \State\Call{ComputeHeatTransfer}{\,}\Comment{Algorithm \ref{lst:compute_heat_transfer}}
      \State\Call{FluidSolve}{\,}\Comment{Algorithm \ref{lst:fluid_solver}}
      \State $t \gets t + \Delta t$
    \EndWhile
    \EndProcedure
  \end{algorithmic}
\end{algorithm}

\subsection{Particle to Grid Transfer}
\label{sec:part_to_grid}

For scientific evaluations it can be quite cumbersome to use particle data, since many visualization techniques, i.e. line plots, contour plots, stream lines, etc. rely on the ability to interpolate the underlying data at any point in space.
While this can be achieved by using the common SPH summation from Equation \eqref{eq:sph_sum}, this method suffers from the issue of particle deficiency at fluid boundaries where the interpolation quality will decrease rapidly.
The ability to reconstruct constant functions can be restored by using normalization, as is also done by Komen et al.~\cite{KST18}, which often yields satisfactory results in practice.
This normalized SPH interpolation is also often called Shepard interpolation in literature and simply normalizes Eq. \eqref{eq:sph_sum}
\begin{equation}
\label{eq:shepard_sum}
A(\bm{x}) = \frac{\sum_{j\in \mathcal{N}_x} V_j A_j W(\bm{x} - \bm{x}_j; h)}{\sum_{j\in \mathcal{N}_x} V_j W(\bm{x} - \bm{x}_j; h)},
\end{equation}
which locally restores the property of accurately representing constant functions.

Nevertheless, we propose a novel method which does not rely on the SPH kernel for interpolation of particle data onto a grid and provides an efficient and accurate method of transferring the underlying continuous field from the particle nodes to specific grid nodes.
Our proposed method overlays a regular grid (in theory this could be any kind of unstructured grid) and solves a linear least squares problem for the values at the grid nodes
\begin{equation}
\label{eq:least-squares}
    \min \sum_{i=1}^N (f(\bm{x}_i; \bm{\theta}) - y_i)^2,
\end{equation}
where $f(\bm{x}_i; \bm{\theta})$ is the function which linearly interpolates the values at the nodes of the grid $\bm{\theta} = (\theta^0, \dots, \theta^M)$ $\in \mathcal{R}^M$ to the position $\bm{x}_i$ of particle $i$ and $y_i$ is the field value at particle $i$. 
The normal equations for this problem are given by
\begin{equation}
    \mathbf{M}^T\mathbf{M}\bm{\theta} = \mathbf{M}^T \bm{y} \quad \mathbf{M}_{ij} = w_{ij}.
\end{equation}
The matrix $\mathbf{M} \in \mathcal{R}^{N \times M}$ has as many rows as there are particles and as many columns as there are grid nodes.
$w_{ij}$ denotes the linear interpolation weight of particle $i$ from grid node $j$.
Matrix $\mathbf{M}$ is very sparse and is solved using the matrix-free conjugate gradient solver of Eigen.
Since there are not always particles in every grid cell, the matrix $\mathbf{M}$ is extended by a set of identity equations for grid nodes where there are no particles in the vicinity and the right hand side is equal to a default value for the interpolated field.

Using this, or other interpolation methods, and depending on the desired grid resolution, it may be possible for gaps to appear in the interpolation grid, where no particles contribute to the interpolated value.
In order to avoid this we perform a smoothing step where the values in the missing cells may be filled in by using Laplacian smoothing
\begin{equation}
    \Delta \bm{\theta} = 0.
\end{equation}
Other smoothing kernels, e.g. Bi-Laplacian or Gaussian kernels, may also be used, but we have found the simple 7 point 3D Laplacian kernel to work well for our purposes.
To do the smoothing step, we implemented a fast marching algorithm to select the grid nodes which are directly adjacent to grid nodes with neighboring particles.
For the smoothing operation, the already computed grid values serve as boundary conditions such that gradients from the already computed fields are preserved and the gaps can be filled in with meaningful values.
The actual boundary of the enclosing grid is treated using zero-gradient boundary conditions.
In practice, we frequently used a grid size equal to the particle diameter and smoothed for up to a distance of 1 to 3 nodes.
This yielded very good results in the regions with gaps as is discussed in the evaluation of Figure \ref{fig:compare_interpolation}.

In comparison to the usual interpolation technique which uses SPH to average local information at a specific position, our method attempts to approximate the underlying field globally and continuously over all particles.
In practice, we have found this to result in smoother interpolated fields while at the same time having equal or better approximation quality.
A brief evaluation of our interpolation method is shown in Section \ref{sec:grid_interpolation_results}.

\subsection{Boundary Conditions in SPH}

Boundary volumes are taken into account using the analytical approach for Volume Maps by Bender et al.~\cite{BKWK19}.
As the name suggests, the method computes analytical volume contributions of solid objects to the SPH integral as a precomputation step, so that it can be evaluated using a simple lookup and polynomial interpolation.
As discussed in the original paper, this method has the advantage of accurately representing solid objects and boundary conditions, and the resulting surface varies smoothly with position.
When the boundary is sampled using particles, e.g. using the method of Akinci et al.~\cite{AIA+12}, particle movement becomes bumpy due to the unevenness of the discretization.

Using Volume Maps, boundaries can simply be integrated into the typical SPH summation of Equation \eqref{eq:sph_sum} resulting in the following equation
\begin{equation}
    A_i = \sum_{j \in \mathcal{N}_i^f} V_j^f A_j^f W_{ij} + V_i^b A_i^b W_{ib},
\end{equation}
where the superscript $f$ denotes contributions from fluid particles and $b$ from boundary objects.
The volume $V_i^b$ denotes the precomputed volume occupied by the boundary in the compact support radius of particle $i$, while the weight $W_{ib}$ is computed for a representative particle on the surface of the boundary.
Finally, $A_i^b$ is chosen to comply with prescribed boundary conditions, e.g. being set to a fixed value for Dirichlet boundary conditions. 
For our model, the Volume Maps boundaries are mainly used in order to extend the fluid domain in order to avoid particle deficiencies for the computation of density and pressure.

We use free slip boundaries at all surfaces, meaning that
\begin{equation}
    \frac{\partial \bm{v}_t}{\partial \bm{n}}\bigg\rvert_{\partial\Omega} = 0,
\end{equation}
where $\partial\Omega$ denotes the entirety of the bounding cylinder (see Figure \ref{fig:calc_region_FEM}), while $\partial / \partial\bm{n}$ denotes the derivative in normal direction.
The velocity in tangential direction to the surface is denoted by $\bm{v}_t$.
This does not require any special treatment in the SPH formulation.
Additionally, we impose adiabatic boundary condition for the heat equation
\begin{equation}
    \frac{\partial T}{\partial \bm{n}}\bigg\rvert_{\partial\Omega} = 0,
\end{equation}
for the entire boundary, while additionally adding a source term at the top surface (line $DC$ in Figure \ref{fig:calc_region_FEM}) of the boundary
\begin{equation}
    \dot{q}''\bigg\rvert_{DC} = \dot{q}''_\text{surface},
\end{equation}
which is computed using the heat source from Equation \eqref{eq:heat_source}.
The adiabatic condition is also satisfied by construction, while the method described in Section \ref{sec:heat_sources} is used to apply the heat sources.

\subsection{2D-FEM Model}

\subsubsection{Model of electromagnetic processes in COMSOL}
\label{sec:EM_Comsol}

For the magnetostatic simulation by means of COMSOL Multiphysics we use the so called $\bm{A}_{m}-V$ formulation which gives equations for the magnetic vector potential $\bm{A}_{m}$ and the scalar electric potential $V$
\begin{equation}
\label{eq:laplace_A}
\Delta \bm{A}_{m}=-\mu_{0}\bm{J},
\end{equation}
\begin{equation}
\label{eq:magn_field}
\bm{B}=\nabla\times\bm{A}_{m},
\end{equation}
\begin{equation}
\label{eq:laplace_phy}
\nabla\cdot(\sigma \nabla V)=0.
\end{equation}
Here $\bm{J}$ and $\bm{B}$ are the current density and magnetic induction respectively, $\mu_0$ is the permeability of vacuum and $\sigma$ is the electrical conductivity. 
The volumetric Lorentz force is governed by
\begin{equation}
\label{eq:lorentz_force}
\bm{F_L}=\bm{J}\times\bm{B}.
\end{equation}
The governing equations, Equation \eqref{eq:laplace_A} and Equation \eqref{eq:laplace_phy}, are subject to the boundary conditions
\begin{equation}
\label{eq:bc_A}
\bm{n}\times\bm{A}\Big |_{AB\bigcup BC\bigcup CD }=0,
\end{equation}
\begin{equation}
\label{eq:bc_V1}
\frac{\partial V}{\partial r}\Big |_{BC}=0,
\end{equation}
\begin{equation}
\label{eq:bc_V2}
V\Big |_{AB}=0,
\end{equation}
\begin{equation}
\label{eq:bc_V3}
\sigma\frac{\partial V}{\partial z}\Big |_{CD}=J_n,
\end{equation}
where $J_n$ denotes the current density distribution along surface $CD$.

\subsubsection{Model of electromagnetic processes in WM}
 \label{sec:EM_WM}
\begin{figure}[ht]
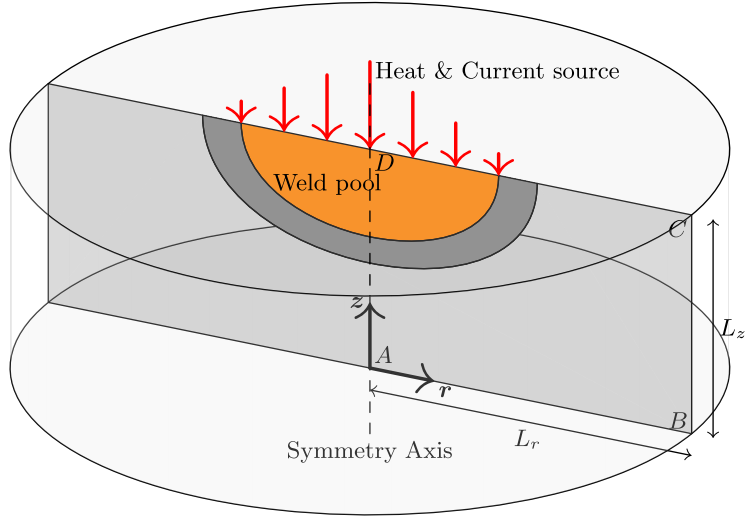

    \centering
\begin{asy}[width=.6\linewidth]
settings.outformat = "pdf";
settings.prc = false;
settings.render = 0;

//size(10cm, 0);
import three;

currentprojection = orthographic(5,5,10, up=Y);
usepackage("bm");
defaultpen(fontsize(9pt));

// Define common paths
path3 weld_pool = path3((0.003, 0.005){S}..{W}(0, 0.003){W}..{N}(-0.003, 0.005));
path3 main_box = (-0.0075, 0, 0)--(-0.0075, 0.005, 0)--(shift((0, -0.005*0.3, 0))*scale3(1.3)*reverse(weld_pool))--(0.0075, 0.005, 0)--(0.0075, 0, 0)--cycle;
path3 weld_pool_solid = shift((0, -0.005*0.3, 0))*scale3(1.3)*weld_pool--reverse(weld_pool)--cycle;
path3 bottom_circle = path3(scale(0.0075)*unitcircle, plane=ZXplane);
path3 top_circle = shift((0, 0.005, 0)) * bottom_circle;
path3 segment  = -0.0075X -- -0.0075X + 0.005Y;

real opac = 0.2;

// Draw bottom circle and surface
draw(surface(bottom_circle), material(grey+opacity(opac)));
draw(bottom_circle);

// Draw surfaces
draw(surface(segment, n=24, c=O, angle1=180, angle2=360, axis=Y), gray+opacity(opac), light=nolight);
draw(surface(main_box), surfacepen=material(darkgrey+opacity(0.2)), light=nolight);
draw(surface(weld_pool_solid), surfacepen=gray, light=nolight);
draw(surface(weld_pool--cycle), surfacepen=orange, light=nolight);

// Draw box outlines
draw(main_box);
draw(weld_pool_solid);
draw(weld_pool--cycle);

// Top circle surface and lines on top
draw(surface(top_circle), material(grey+opacity(opac)));
draw(top_circle);

// Heat arrows
for (int i=-3; i<=3; ++i){
    draw(shift(i*0.001, 0, 0) * ((0, 0.007 - 0.0005*abs(i), 0) -- (0, 0.005, 0)), p=linewidth(1.3)+red, arrow=Arrow3(TeXHead3));
}

// Draw axes etc
draw((0,-0.0015, 0)--(0,0.0065, 0), dashed);
draw((0, 0, 0)--(0,0.0015, 0), linewidth(1.3), arrow=Arrow3(TeXHead3));
draw((0, 0, 0)--(0.0015,0, 0), linewidth(1.3), arrow=Arrow3(TeXHead3));
draw("$L_z$", (0.008, 0, 0)--(0.008, 0.005, 0), arrow=Arrows3(TeXHead3));
draw("$L_r$", (0, -0.0005, 0)--(0.0075,-0.0005, 0), arrow=Arrows3(TeXHead3));

// Dont know how to draw bars in 3d, somehow ambiguous
// draw((0, -0.0005, 0)--(0.0075,-0.0005, 0), bars=Bars3);
// draw((0.008, 0)--(0.008, 0.005), bars=Bars3);

// Labels on top of almost everything
label("$A$", (0, 0, 0), NE);
label("$B$", (0.0075, 0, 0), NW);
label("$C$", (0.0075, 0.005, 0), SW);
label("$D$", (0, 0.005, 0), SE);
label("Symmetry Axis", (0, -0.0015, 0), S);
label("$\bm{z}$", (0,0.0015, 0), W);
label("$\bm{r}$", (0.0015,0, 0), SE);
label("Heat \& Current source", (0.00002, 0.0065, 0), NE);
label("Weld pool", (-0.001, 0.004, 0));

// Front side of the cylinder in front of everything
draw(surface(segment, n=24, c=O, angle1=0, angle2=180, axis=Y), gray+opacity(opac), light=nolight);

\end{asy}
    \caption{Computational domain for both SPH and FEM simulations. The FEM methods solve the problem in rotational symmetry in the $ABCD$ plane, while SPH considers the full cylinder by explicit rotation of $ABCD$ around the denoted symmetry axis.}
    \label{fig:calc_region_FEM}
\end{figure}

The model of electromagnetic processes is based on the stationary magnetic field diffusion equation \cite{davidson2001} which in the case of axial symmetry reads as follows
\begin{equation}\label{eq:magn_diffusion}
\frac{1}{r}\frac{\partial }{\partial r}\left(r\frac{\partial B_{\theta}}{\partial r}\right)+\frac{\partial^2 B_{\theta}}{\partial z^2}-\frac{B_{\theta}}{r^2}=0,
\end{equation}
where $B_{\theta}$ is an azimuthal component of magnetic induction. From Ampere's law 
\begin{equation}
\label{eq:amperes_law}
\mu_0\bm{J}=\nabla\times \bm{B},
\end{equation}
the next boundary conditions for Equation \eqref{eq:magn_diffusion} can be readily obtained
\begin{equation}
\label{eq:bc_magn_diff_1}
B_{\theta}\Big |_{CD}=\frac{\mu_0}{r}\int_{0}^{r} J_n(s)sds,
\end{equation}
\begin{equation}\label{eq:bc_magn_diff_3}
B_{\theta}\Big|_{BC}=\frac{\mu_0I_w}{2\pi L_r},
\end{equation}
\begin{equation}\label{eq:bc_magn_diff_2}
\frac{\partial B_{\theta}}{\partial z}\Big|_{AB}=0.
\end{equation}
Here $L_r$ is the radius of computational domain (see Figure \ref{fig:calc_region_FEM}), while $I_w$ denotes the welding current.
Under the given magnetic field, the current density and Lorentz force are obtained from Equation \eqref{eq:amperes_law} and Equation \eqref{eq:lorentz_force}, respectively. 
 
\subsubsection{Numerical procedure in COMSOL Multiphysics}
 \label{sec:numerical_procedure_comsol}

\begin{figure}
    \centering
    \def\svgwidth{.5\textwidth}
    \includegraphics[width=.7\linewidth]{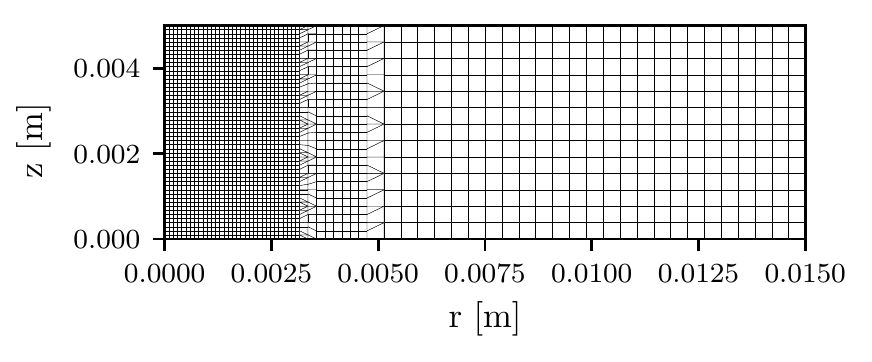}
    \caption{COMSOL mesh as mixture of quadrilaterals and triangles.}
    \label{fig:COMSOL_mesh}
\end{figure}

The governing equations in COMSOL are solved using the Finite Element Method (FEM). For temporal discretization the BDF (backward differentiation formula) was applied, with default settings of COMSOL 5.6. The spatial discretization can be seen in Figure \ref{fig:COMSOL_mesh}. Here the nodes used quadratic shape functions for temperature, velocity, magnetic strength and for scalar and vector potential. The shape functions utilized for the solution of the pressure were linear. The sparse direct solver PARDISO was selected to solve the resulting system of linear equations.

\subsubsection{Numerical procedure in WM}
\label{sec:numerical_procedure_WM}
\begin{figure}
    \centering
    \def\svgwidth{.5\textwidth}
    \includegraphics[width=.7\linewidth]{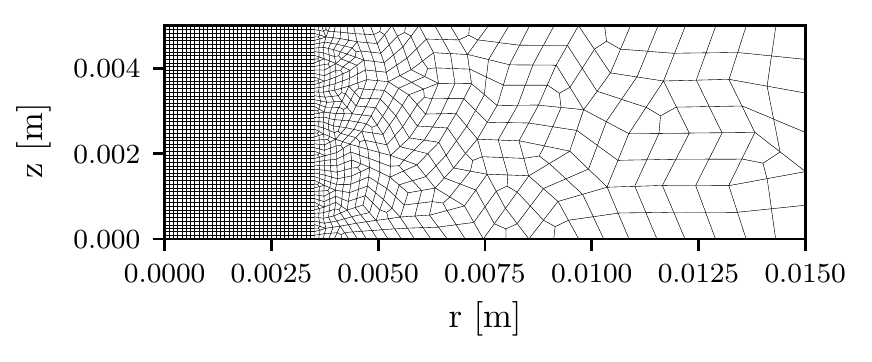}
    \caption{FEM mesh as Quad9 quadrilaterals. The Navier-Stokes equations are solved in the region of interest $\Omega=\{ \SI{0}{\milli\meter}<r< \SI{3.5}{\milli\meter}, \SI{0}{\milli\meter}<z< \SI{5}{\milli\meter}\}$ while the heat transport is analyzed in whole domain.}
    \label{fig:FEM_mesh}
\end{figure}

The governing equations are solved using the Galerkin Finite Element Method (FEM) \cite{zienkiewicz2014}. 
A characteristic-based scheme \cite{zienkiewicz1995} was used in temporal discretization of Equation \eqref{eq:navier_stokes} and Equation \eqref{eq:energy_transport} along with FEM approximation in space, see Figure \ref{fig:FEM_mesh}. 
Nine node biquadratic shape functions are employed for velocity, temperature and magnetic fields whereas the pressure is approximated by four node bilinear shape functions. 
Such a choice of mixed interpolation for velocity and pressure fields provides stability of the numerical procedure \cite{zienkiewicz2014}. 
The sparse direct solver PARDISO \cite{schenk2004} was employed to solve the resulting systems of linear equations. All the numerical algorithms were implemented in the Wolfram Mathematica Language (WM).   

\section{Results}
\label{sec:results}
\subsection{Grid Interpolation}
\label{sec:grid_interpolation_results}
As noted before, while all following evaluations could technically be done on particle data alone, many existing evaluation functions and programs, e.g. ParaView and Tecplot, are tailored to grid structures and unstructured meshes.
This makes the usage of regular and unstructured meshes very appealing for quantitatively evaluating and comparing scientific computations.
To this end, we compare the interpolation accuracy using na\"ive SPH interpolation (see Equation \eqref{eq:sph_sum}), Shepard corrected SPH interpolation (see Equation \eqref{eq:shepard_sum}) and our proposed least-squares linear interpolation (see Equation \eqref{eq:least-squares}).

\begin{figure}
    \centering
    \includegraphics[width=.5\linewidth]{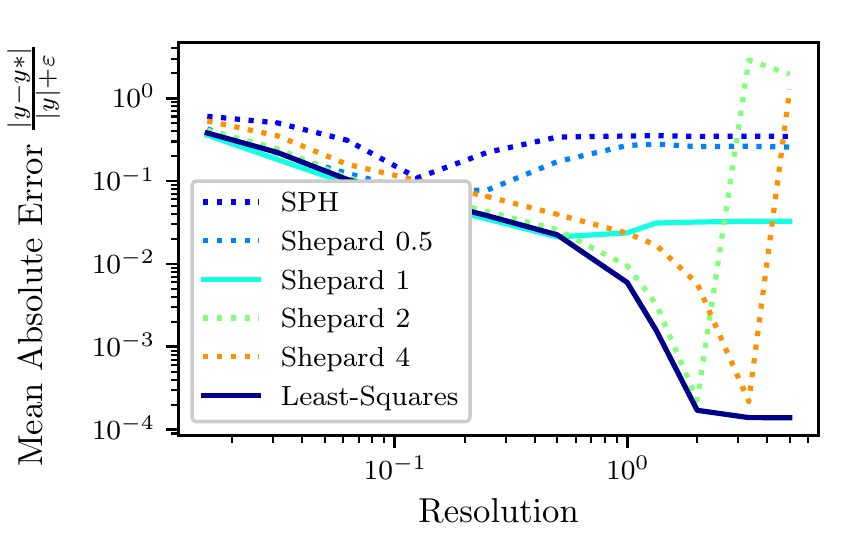}
    \caption{Comparison of particle to grid interpolation techniques.}
    \label{fig:particle_to_grid_interpolation}
\end{figure}

Figure \ref{fig:particle_to_grid_interpolation} shows a comparison of the different interpolation approaches.
Each method has been used to compute the particle to grid transfer of the same particle data. 
The error is computed by interpolating these grids back to the original particle positions and computing the mean absolute error.
The error is determined by evaluating the error at the particle positions using linear interpolation from the computed grids.
The underlying particle data is the velocity field of a randomly selected frame of a fluid dynamics simulation in motion, but the same trend is visible regardless of which field is approximated.
The regular grid spacing $d_{spacing}$ is determined by a multiple $n_{grid}$ of the SPH particle radius $r_{part}$ 
\begin{equation}
d_{spacing} = n_{grid} \cdot r_{part}.
\end{equation}
The resolution of the grid shown in Figure \ref{fig:particle_to_grid_interpolation} is then computed as the inverse of this factor, $n_{grid}^{-1}$, i.e. a higher resolution corresponds to a smaller grid spacing $d_{spacing}$ and thus a larger number of total grid nodes.

The Shepard interpolation was evaluated using a varying factor for the SPH compact support radius depending on the factor for the grid spacing  $h_{shepard} = d_{spacing} \cdot n_{shepard}$.
It can be seen that both our proposed least-squares interpolation as well as Shepard interpolation using the grid spacing with a factor of 1 and 2 as compact support radius perform the best in this comparison, for grid spacings larger than $r_{part}$.
Na\"ive SPH interpolation performs best when the compact support radius is equal to the compact support used in the actual simulation.
It also suffers from worse interpolation quality due to particle deficiency at the surface, leading to large errors in the interpolation.

For finer resolutions it can be seen that the least-squares interpolation performs best in all cases, while the Shepard interpolation with a factor of 2 and subsequently a factor of 4 have a comparable error when a grid spacing of $0.5 r_{part}$ and $0.25 r_{part}$ is used respectively.
From this comparison we conclude that both Shepard interpolation, as well as our proposed least-squares interpolation can work very well for particle data.
The interpolation quality of our least-squares approach becomes more apparent when the grid-resolution increases as well.

\begin{figure}
    \centering
    \includegraphics[width=.5\linewidth]{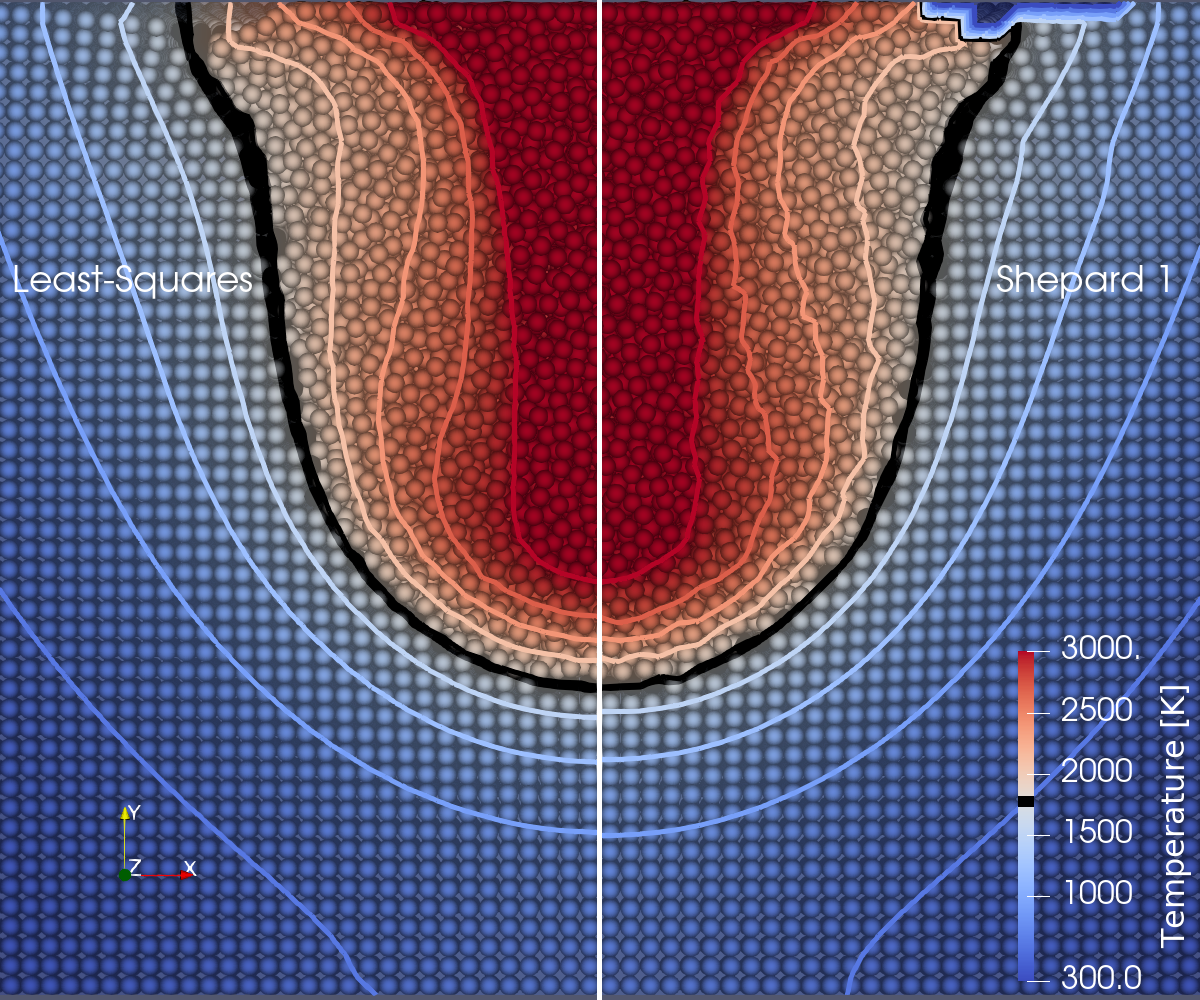}
    \caption{A comparison of our proposed least-squares interpolation method using smoothing of one surrounding grid node (left) and Shepard interpolation (right) without smoothing.}
    \label{fig:compare_interpolation}
\end{figure}

Another advantage of our interpolation can be seen in Figure \ref{fig:compare_interpolation}, where the particle to grid transfer of the temperature field using two different interpolation methods is shown. 
The left half shows the interpolation of our proposed least-squares method and smoothing for up to one neighboring grid node, and the right side shows Shepard interpolation.

From this qualitative comparison it is possible to see that while the values are almost identical in the entire domain, the least squares interpolation results in a much smoother field.
This is especially visible at the temperature contour lines within the black border.
In this figure the advantages of our smoothing approach can also be observed.
While the top right of the Shepard interpolation shows gaps in the interpolation due to missing particles, the left side shows that this part (top left) can be filled in with very plausible values using the aforementioned Laplacian smoothing.
While the filled in values should be treated with care, this kind of filling in allows the consistent reconstruction of fixed domains from particle data.

For all following particle to grid interpolations we use our least-squares interpolation with a grid resolution of twice the particle radius, because this is the same resolution as for the WM and COMSOL simulations.
In addition, we apply a smoothing of up to 3 particle nodes, although only for times close to the end of the simulation is the full smoothing length actually needed.
At this resolution it can be expected to incur a mean absolute error in the interpolated data of 4\% for scalar values.

\subsection{Simulation parameters}

To allow for a first validation, the relevant material parameters have been simplified to constants, see Table \ref{tab:material_parameters}. It should be mentioned that the morphological constant $C$ is usually given without division over the density, where it would evaluate as $C=$\SI{3e8}{\kg \per \m\cubed \per\s}.

\begin{table}[ht]
        \centering
        \begin{tabular}{lll}
        \hline\noalign{\smallskip}
        Property                & Unit      & Value        \\ \noalign{\smallskip}\hline\noalign{\smallskip}
         Density $\rho$ & \si{\kilo \gram \per \metre \cubed} & $8100$ \\
         Surface tension $\gamma$ & \si{\newton \per \metre}& disabled  \\
         Viscosity $\mu$  & \si{\pascal \second}&   0.004 \\
         Thermal conductivity $\lambda$  & \si{\watt \per \metre \per \kelvin} & 22.9 \\
         Electrical conductivity $\sigma$ & S m$^{-1}$ & $10^6$ \\
         Heat capacity $c_{p}$ & \si{\joule\per\kilo\gram\per\kelvin} & 800\\
         Melting temperature $T_{l}$ & \si{\kelvin} & $1773$ \\
         $C$ from Equation \eqref{eq:porosity} & \si{\per \s} & $3.7\cdot10^{4}$\\
        \noalign{\smallskip}\hline 
    \end{tabular}
    \smallskip
        \caption{Material parameters}
\label{tab:material_parameters}
\end{table}

The properties are applied homogeneously over the calculation domain, which consists of a rotationally symmetrical cylinder, as seen in Figure \ref{fig:calc_region_FEM}. The details of the computational domain can be seen in Table \ref{tab:domain_parameters}. It is important to note that the SPH domain is realized in full 3D, while the FEM domain is just considering a 2D plane with rotational symmetry around the symmetry axis AD.

\begin{table}[h]
        \centering
        \begin{tabular}{lll}
        \hline\noalign{\smallskip}
        Property                & Unit      & Value        \\ \noalign{\smallskip}\hline\noalign{\smallskip}
         Radius of the cylinder $L_{r}$ & m & $15 \cdot 10^{-3}$ \\
         Height of the cylinder $L_{z}$ & m & $5 \cdot 10^{-3}$ \\
         Initial temperature & K & 300\\
         Heat source center \bm{$x_0$} & m & $(0,\,6 \cdot 10^{-3},\,0)$\\
         Heat source dist. radius $r_{HS}$ & m & $2.7 \cdot 10^{-3}$\\
         Heat source power $P_{source}$ & W & 1213.8 \\
         Total current $I_{w}$ & \SI{}{\ampere} & 140 \\
         External forces $\bm{f}_{ext}$ & N m$^{-3}$ & See Figure \ref{fig:EM_forces}\\

         \noalign{\smallskip}\hline 
    \end{tabular}
    \smallskip
    \caption{Domain parameters}
    \label{tab:domain_parameters}
\end{table}

\begin{figure}
     \centering
     \includegraphics[width=0.5\linewidth]{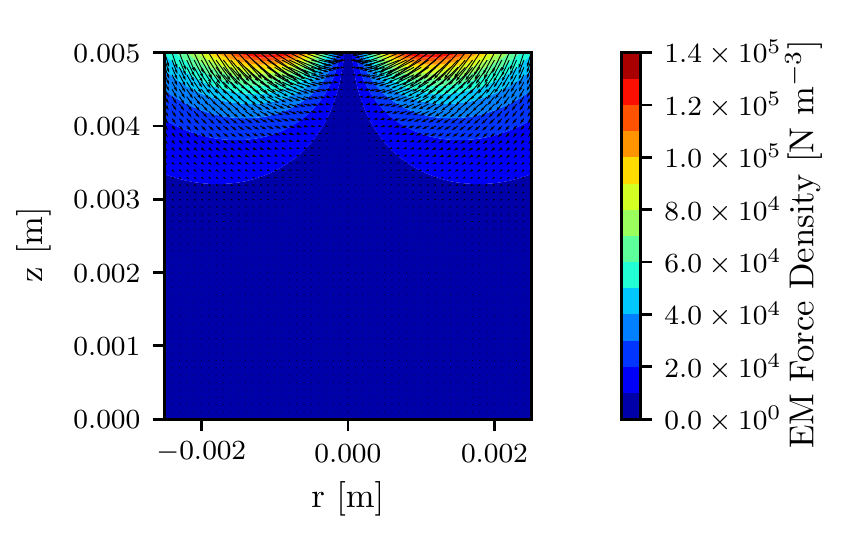}
     \caption{Lorentz force in the region of interest with the direction indicated as an arrow and the magnitude indicated as the underlying color.}
     \label{fig:EM_forces}
 \end{figure}

The external Lorentz force field due to electromagnetic forces is pre-computed in advance of the fluid-flow simulation, according to Equation \eqref{eq:lorentz_force} in order to be able to use the same external forces for both the SPH and WM simulations. 
The magnetic strength field is obtained from the solution of Equation \eqref{eq:magn_diffusion} by means of the FEM. From these the Lorentz forces $\bm{F}_L$ are computed by Equation \eqref{eq:lorentz_force}, and are then included as external volumetric forces $\bm{f}_{ext}$ in the Navier-Stokes equation Equation \eqref{eq:navier_stokes}, by a rotation of the 2D plane around the symmetry axis
\begin{equation}
\bm{f}_{ext} = \bm{F_L}.
\label{eq:external_forces}
\end{equation}
The vector field of the forces is calculated from the applied electric current density $\bm{J_n}(s)$ in Equation \eqref{eq:bc_magn_diff_1}, distributed as a Gaussian along the DC boundary (see Figure \ref{fig:calc_region_FEM}) in the same way that the heat source distribution is defined in Equation \eqref{eq:heat_source}, but with the welding current $I_{w}$
\begin{equation}
\bm{J_n}(s) = I_{\text{w}} \frac{3}{\pi r_{HS}^2}\exp\left(-3\frac{||\bm{x_i} - \bm{x_0}||^2}{r_{HS}^2}\right).
\end{equation}
Figure \ref{fig:EM_forces} shows the Lorentz force in the region of interest, that was pre-calculated with the WM-code, according to Section \ref{sec:EM_WM} as the electrical conductivity was constant. The resulting force field was then applied in rotational symmetry on the SPH particles. The Lorentz force for COMSOL were calculated within the software according to Section \ref{sec:EM_Comsol} and they were found to be nearly identical in the region of interest, showing only small deviations far away due to the difference of the system of equations (c.f. Section \ref{sec:EM_Comsol} and Section \ref{sec:EM_WM}).

The convergence of the simulation was achieved using the numerical parameters listed in Table \ref{tab:numerical_parameters_SPH} and Table \ref{tab:numerical_parameters_FEM}. 
The SPH particle diameter is chosen to be equivalent to the element size in the region of interest for the Wolfram Mathematica and COMSOL FEM simulation.
The WM method and the SPH method use adaptive time integration using the CFL criterion, with identical maximum time step sizes but different factors.
The WM method made use of both second-order consistent implicit time integration as well as higher order finite-element basis functions.
As such, the CFL factor in SPH was chosen heuristically to be smaller, since there a only first-order implicit time integration scheme with operator splitting is used.
For COMSOL, the backwards differentiation formula (BDF) was used for temporal discretization, with default parameters.

\begin{table}
    \centering
    \begin{tabular}{lll}
    \hline\noalign{\smallskip}
        Property                & Unit      & Value        \\ \noalign{\smallskip}\hline\noalign{\smallskip}
         SPH particle radius & m & $5\cdot10^{-5}$\\
         Maximum time step & s & $1\cdot10^{-4}$\\
         CFL criterion & - & $0.1$\\
        \noalign{\smallskip}\hline 
    \end{tabular}
    \smallskip
    \caption{Numerical parameters of SPH}
    \label{tab:numerical_parameters_SPH}
\end{table}

\begin{table}
    \centering
    \begin{tabular}{lll}
    \hline\noalign{\smallskip}
        Property                & Unit      & Value        \\ \noalign{\smallskip}\hline\noalign{\smallskip}
         Element size in region of interest & m & $10 \cdot 10^{-5}$\\
         Maximum time step & s & $1\cdot10^{-4}$\\
         CFL criterion & - & $0.5$\\
        \noalign{\smallskip}\hline 
    \end{tabular}
    \smallskip
    \caption{Numerical parameters of WM}
    \label{tab:numerical_parameters_FEM}
\end{table}

\subsection{Simulation results}

\subsubsection{Conductive Heat Transfer}

\begin{figure}
     \centering
     \includegraphics[width=.5\linewidth]{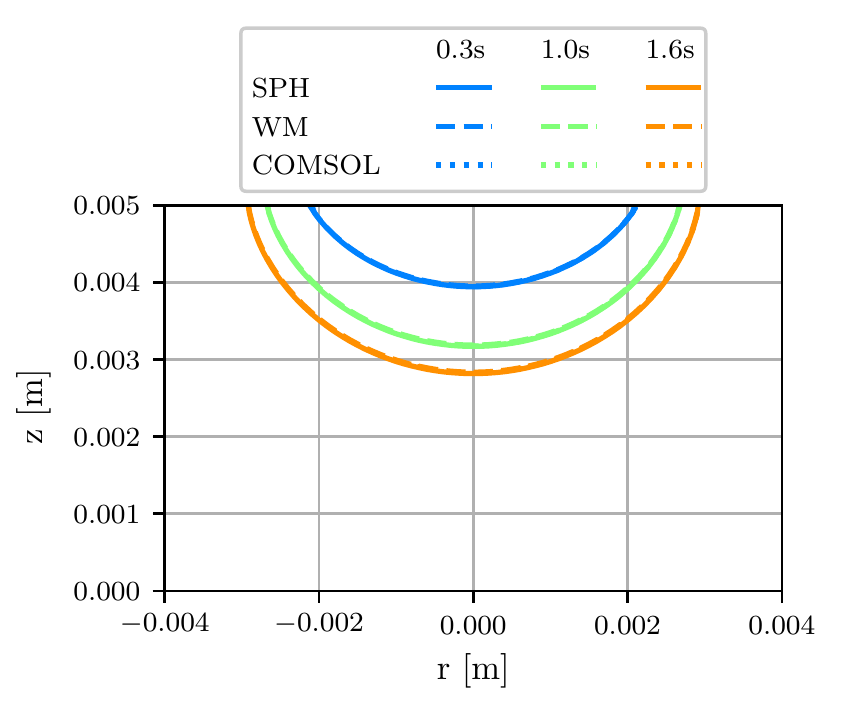}
     \caption{Melting temperature (1773K) iso-contours at different times for SPH,  WM and COMSOL without Lorentz force $\bm{f}_{ext}$=\SI{0}{\N\per\m\cubed}, i.e pure heat conduction.}
     \label{fig:contour_noEM}
 \end{figure}

Before the quantitative comparison of forced convective and conductive heat transfer is performed, a comparison without the external forces $\bm{f}_{ext}$ was made, i.e. without any fluid movement and therefore no convection. As can be seen from the shape of the melting temperature iso-contours in Figure \ref{fig:contour_noEM}, the calculation for the conductive heat transfer was in perfect agreement between SPH, WM and COMSOL for all time steps. Therefore, we consider the purely conductive part of the heat transfer module in SPH as validated.

\subsubsection{Forced Convective and Conductive Heat Transfer}

\paragraph{Quantitative Comparison}

\begin{figure}
    \centering
    \includegraphics[width=.5\linewidth]{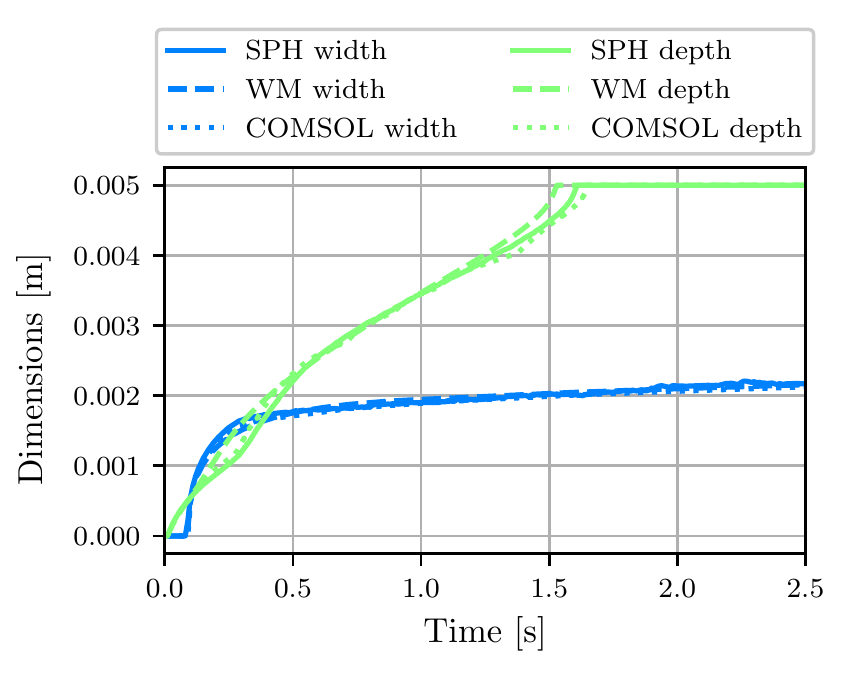}
    \caption{Evolution over time of the melt-pool dimensions depth and width at 4.5mm.}
    \label{fig:pool_dimensions}
\end{figure}

The dimensions of the melt-pool over the course of the simulations are shown in Figure \ref{fig:pool_dimensions}, for the case when the external forces were set to the Lorentz force as shown in Equation \eqref{eq:external_forces}. 
The weld pool dimensions are computed by solving a non-linear algebraic equation for the position of the melting front.
Overall both characteristic melt-pool dimensions, the depth and the width, show very good agreement between all three methods and exhibit a similar development over time. 
The depth of the weld pool for the SPH and COMSOL simulation match almost perfectly across all time steps. 
During the first $t<0.5$\si{\s} the FEM weld pool is deeper, but afterwards agreement between all three methods is again observed for $t>0.5$\si{\s}.
Starting at roughly $t>1.25$\si{\s} the development of the depth increases for the WM solver, reaching full penetration about $\Delta t \sim 0.1$\si{\s} earlier than the SPH method, while the COMSOL solution is a bit slower, reaching full penetration $\Delta t \sim 0.03$\si{\s} later. Also the depth of the COMSOL solution does not seem to develop entirely smoothly, displaying minor kinks at $t=1.35$\si{\s} and $t=1.5$\si{\s}.

As the interpolation of the SPH solution into a grid has limitations close to the boundaries, since fewer particles were present to interpolate the results, the width was not computed exactly on the boundary, but $0.5$\si{\mm} below the surface. 
While we use our proposed interpolation method to obtain a plausible filling in of the gap, a better result is still obtained when looking at the weld pool width a short distance into the domain. 
Here the width matches up nearly perfectly between all three solver methods, with a minor deviation for $t<0.5$\si{\s}, where the COMSOL solution appears right between the SPH and WM solution. 

\begin{figure}[ht]
     \centering
     \includegraphics[width=.5\linewidth]{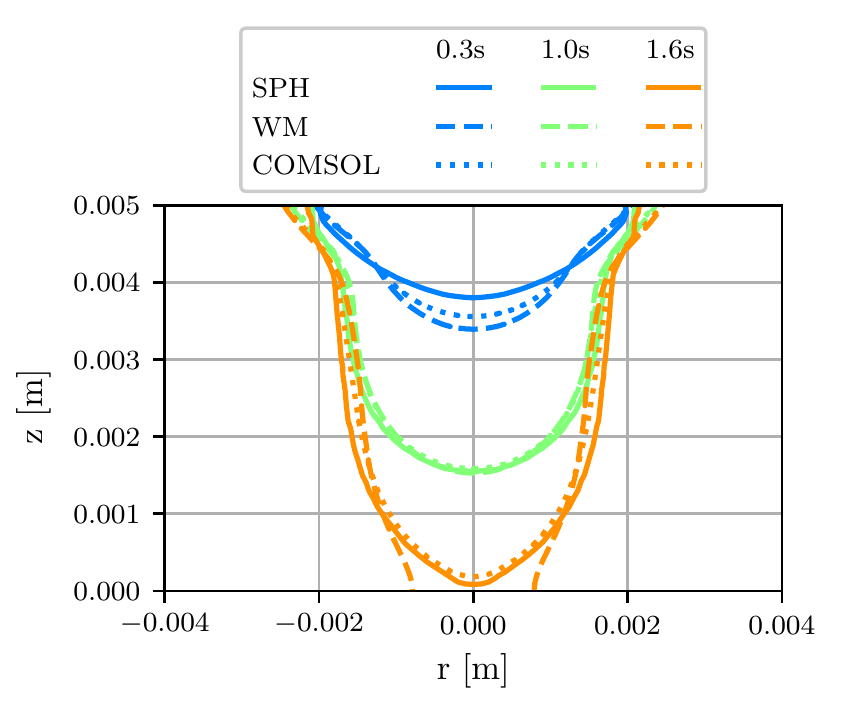}
     \caption{Melting temperature (1773K) iso-contours at different times for SPH and WM, including conductive heat transfer driven by the Lorentz force.}
     \label{fig:temperature_contours}
\end{figure}

In Figure \ref{fig:temperature_contours} the melting temperature iso-contour can be seen, at characteristic time steps of the process for these developments of $0.3$\si{\s}, $1.0$\si{\s} and $1.6$\si{\s}. 
These characteristic times were also chosen for all future comparisons.
At $0.3$\si{\s} the melt-pool is more shallow in SPH, compared to WM and COMSOL, with its melting contour having a very similar shape to that shown in Figure \ref{fig:contour_noEM} for the same time.
This suggests that the convective heat transfer has not yet started in the SPH case, which could be due to insufficient amount of particles to allow for any fluid movement. 
At $1.0$\si{\s} we can see excellent agreement between the SPH, WM and COMSOL method, not just in terms of maximal width or depth but also in the general shape of the melt-pool. 
At $1.6$\si{\s} the agreement between SPH and COMSOL can still be regarded as very good, while WM deviates for these solutions in the lower part of the melt-pool $z \leq 0.001$m, as full penetration of the work piece is already achieved for WM at this time step, whereas both SPH and COMSOL are just about to reach the bottom boundary.
For the intermediate part $0.001$m $\leq z \leq 0.0045$m, the COMSOL solution is right in between the WM and SPH solution. 
It should be mentioned that there are slight deviations in the top part $z > 0.0045$m for the SPH solution, which occur due to missing particles.

\begin{figure}
    \centering
    \includegraphics[width=.5\linewidth]{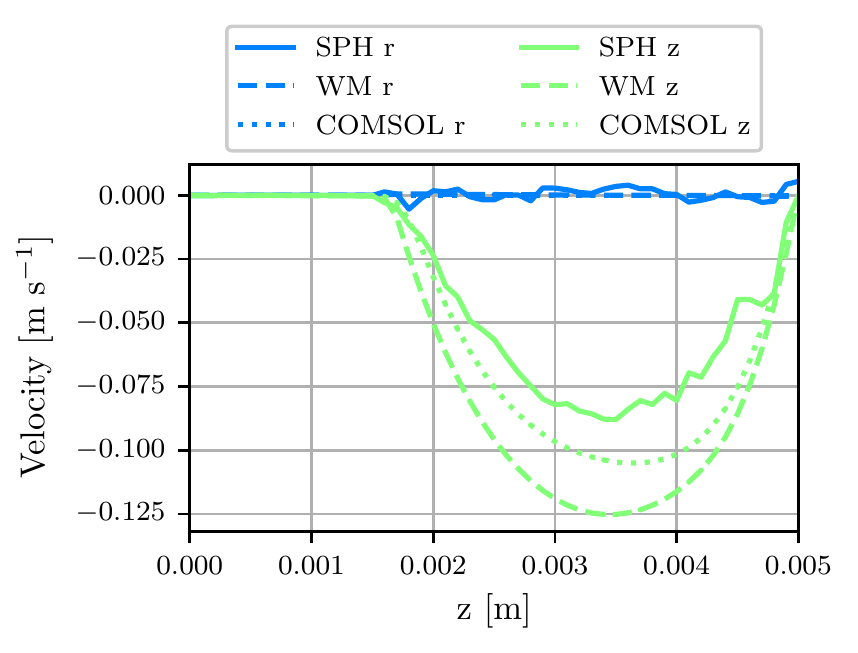}
    \caption{Velocity by component along the symmetry axis at $t=1.0$\si{\s}.}
    \label{fig:vel_height_axis}
\end{figure}

\begin{figure}
    \centering
    \includegraphics[width=.5\linewidth]{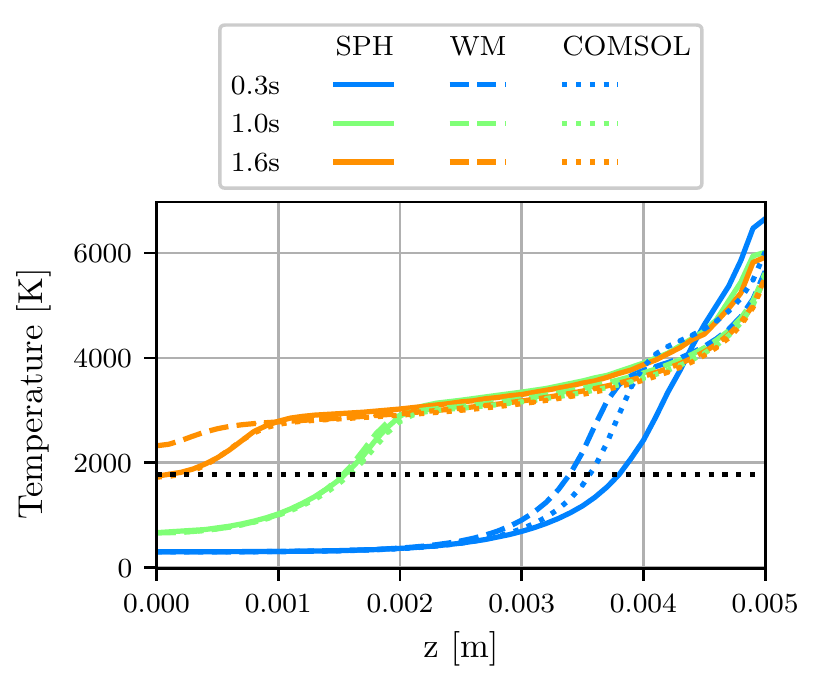}
    \caption{Temperature along symmetry axis at different times. The dotted black line marks the melting temperature (1773K).}
    \label{fig:temp_height_axis}
\end{figure}

To obtain a better understanding of the significant velocities, 
the velocity distributions at $t=1.0$\si{\s} along the symmetry axis (z) are shown in Figure \ref{fig:vel_height_axis}.
The shape of the velocity distribution is very similar for all three solutions, while the difference between WM and COMSOL is quite comparable to the difference between COMSOL and SPH. 
This is especially remarkable since both WM and COMSOL are FEM solvers of the same problem.
In the light of the observed differences between both FEM approaches, the disagreement of the SPH method with the FEM solvers appear to actually be within the margin of error that can be expected from the application of a FEM in general. 
On the contrary, the overall level of agreement is especially remarkable when considering that in the SPH method an entirely different discretization is used and that a whole three dimensional domain is simulated.
The agreement in the radial direction is very good for all three methods. 
For SPH the velocities fluctuate around \SI{0}{\m\per\s}, due to the three dimensional nature of the simulation.
The fluctuations of the velocity for SPH may also be partly attributed to the fact that only a slice of full 3-D data is examined, where no constraints explicitly enforce strict movement within the radial plane.
As such, it is only to be expected for a certain amount of statistical noise to be present in the SPH velocity components, compared to the FEM methods where the considered axis was realized as a rotational symmetry boundary, and as such does not permit movement in radial direction.

\begin{figure*}
\centering
 \begin{subfigure}[t]{.37\textwidth}
    \centering
    \includegraphics[width=1\linewidth]{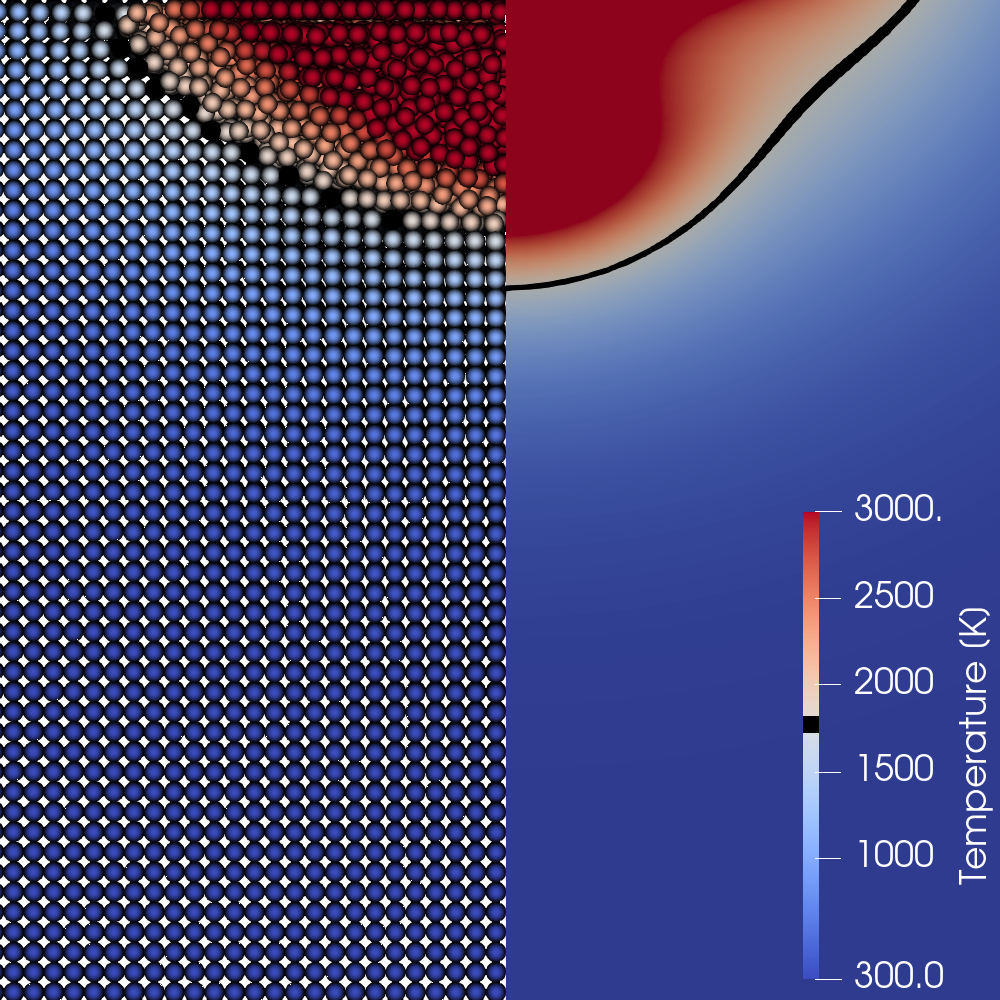}
    \caption{Temperature at t=0.3s.}
    \label{fig:crosssection_temperature_0.3}
\end{subfigure}%
\begin{subfigure}[t]{.37\textwidth}
    \centering
    \includegraphics[width=1\linewidth]{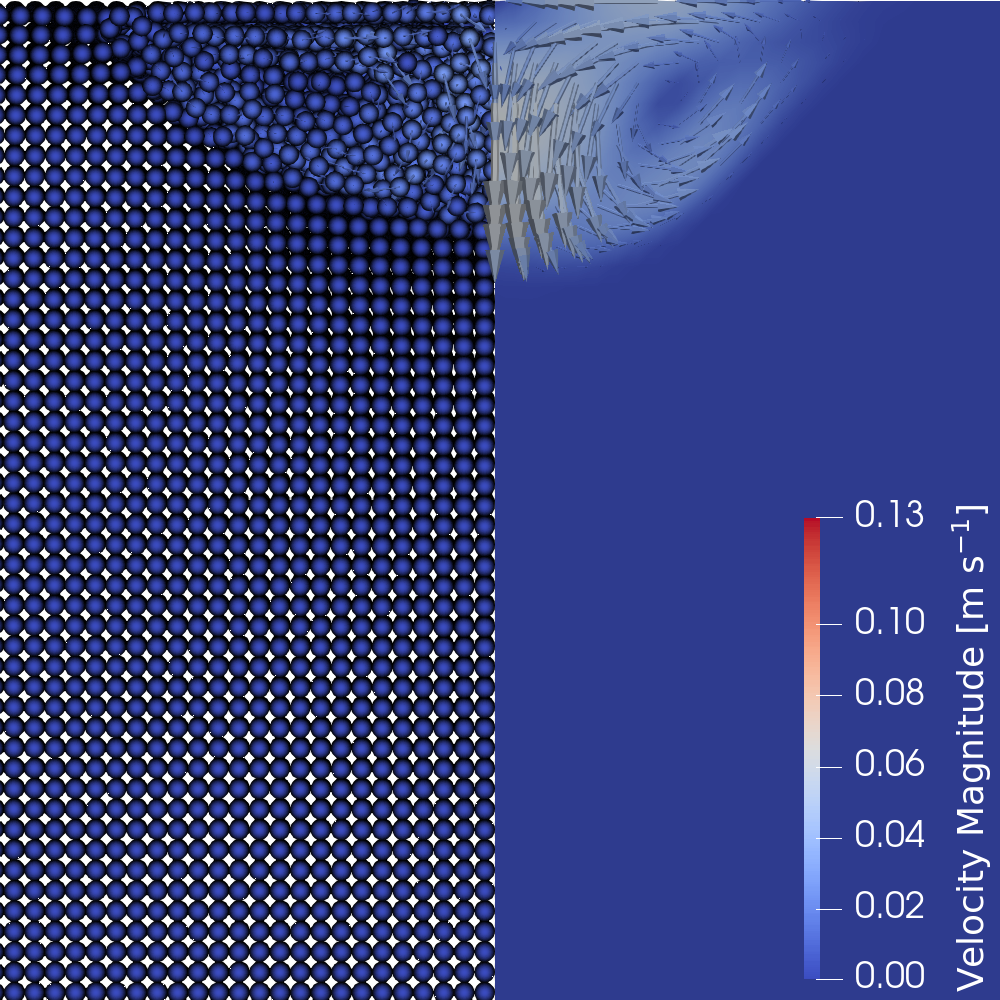}
    \caption{Velocity at t=0.3s.}
    \label{fig:crosssection_velocity_0.3}
\end{subfigure}\\
\begin{subfigure}[t]{.37\textwidth}
    \centering
    \includegraphics[width=1\linewidth]{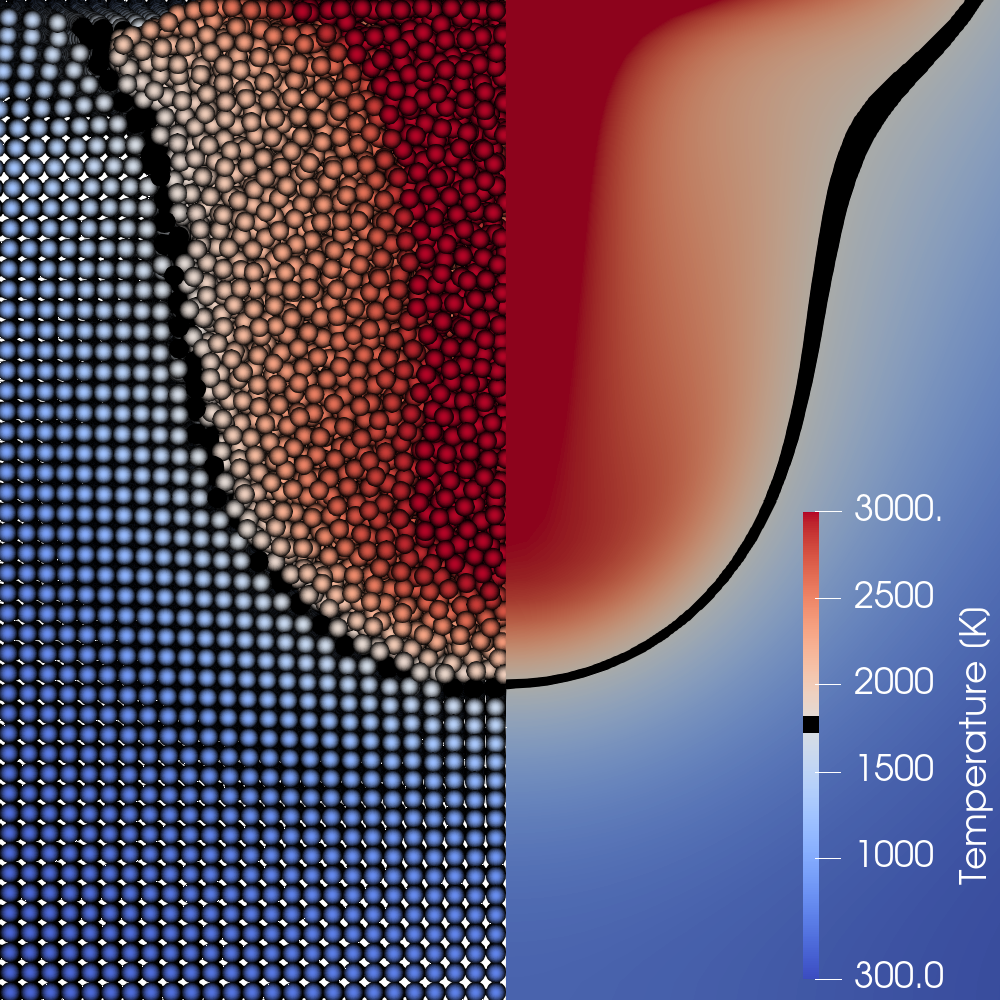}
    \caption{Temperature at t=1s. 
    }
    \label{fig:crosssection_temperature_1.0}
\end{subfigure}%
\begin{subfigure}[t]{.37\textwidth}
    \centering
    \includegraphics[width=1\linewidth]{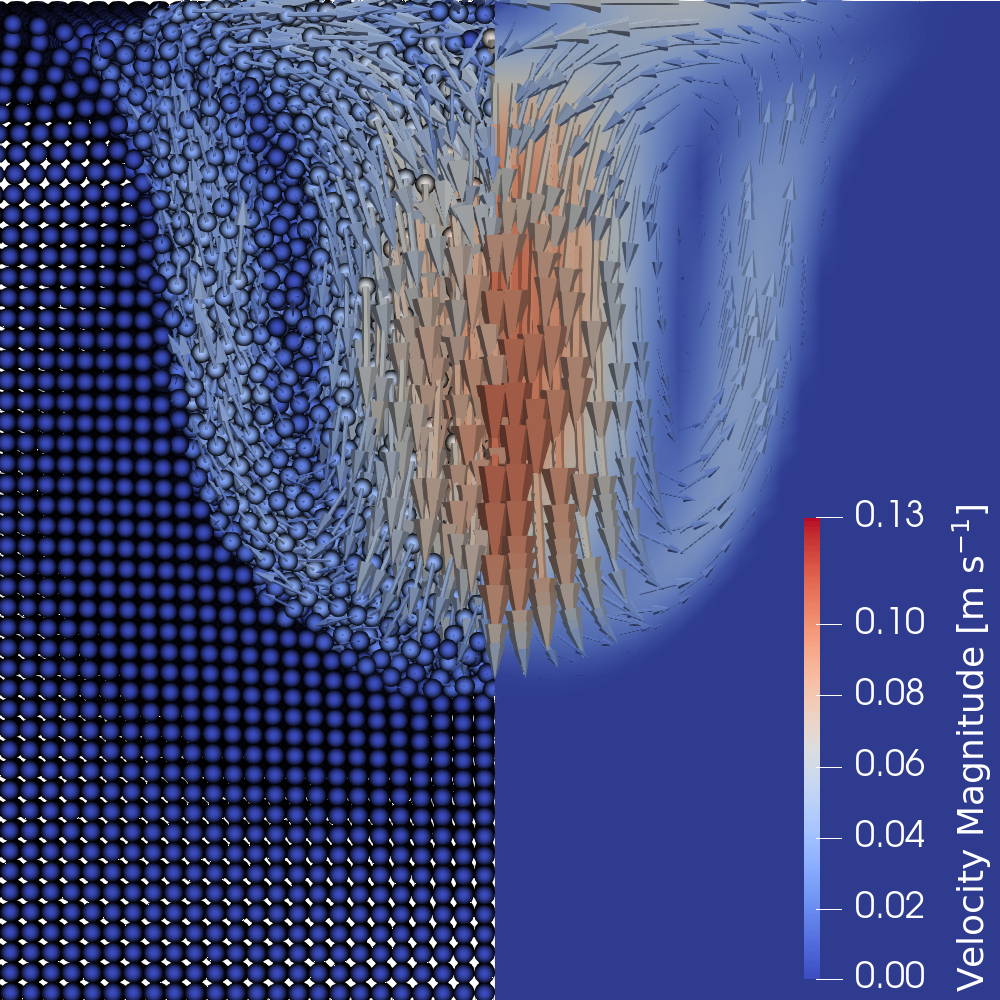}
    \caption{Velocity at t=1s. 
    }
    \label{fig:crosssection_velocity_1.0}
\end{subfigure}\\
\begin{subfigure}[t]{.37\textwidth}
    \centering
    \includegraphics[width=1\linewidth]{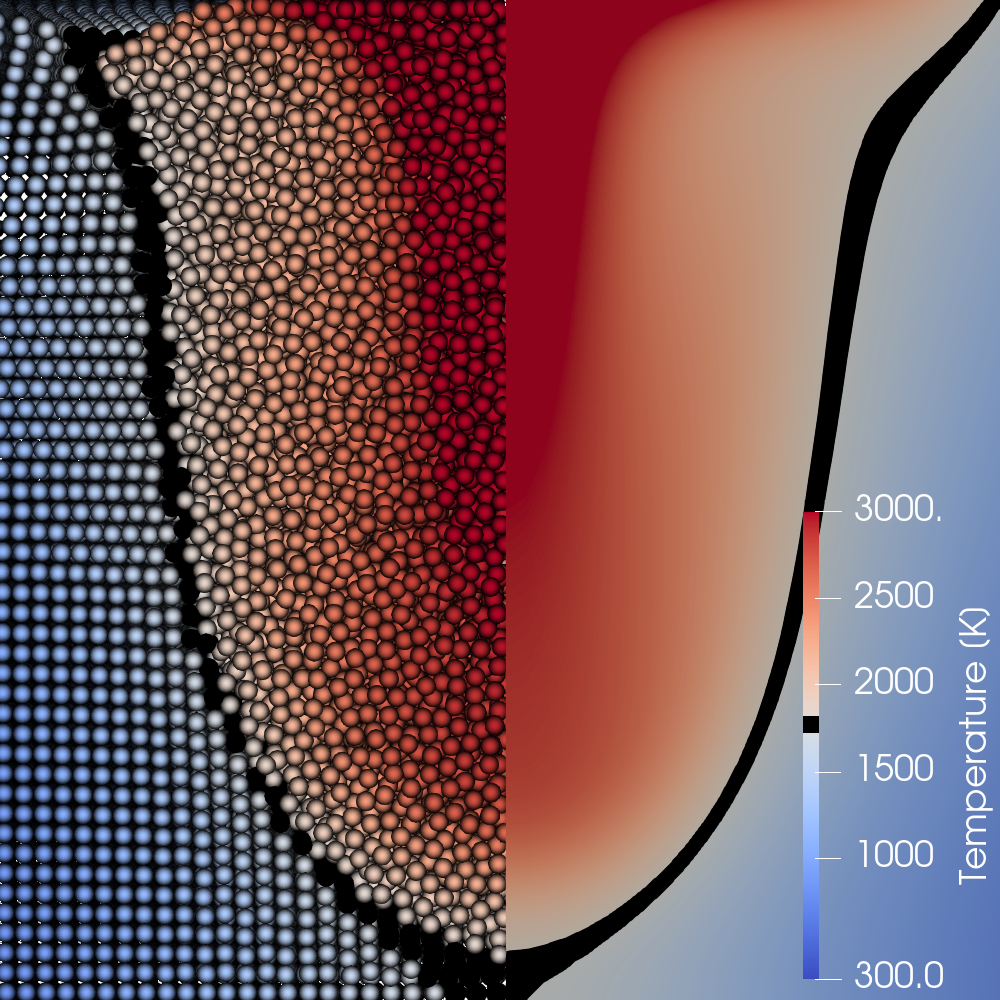}
    \caption{Temperature at t=1.6s. }
    \label{fig:crosssection_temperature_1.6}
\end{subfigure}%
\begin{subfigure}[t]{.37\textwidth}
    \centering
    \includegraphics[width=1\linewidth]{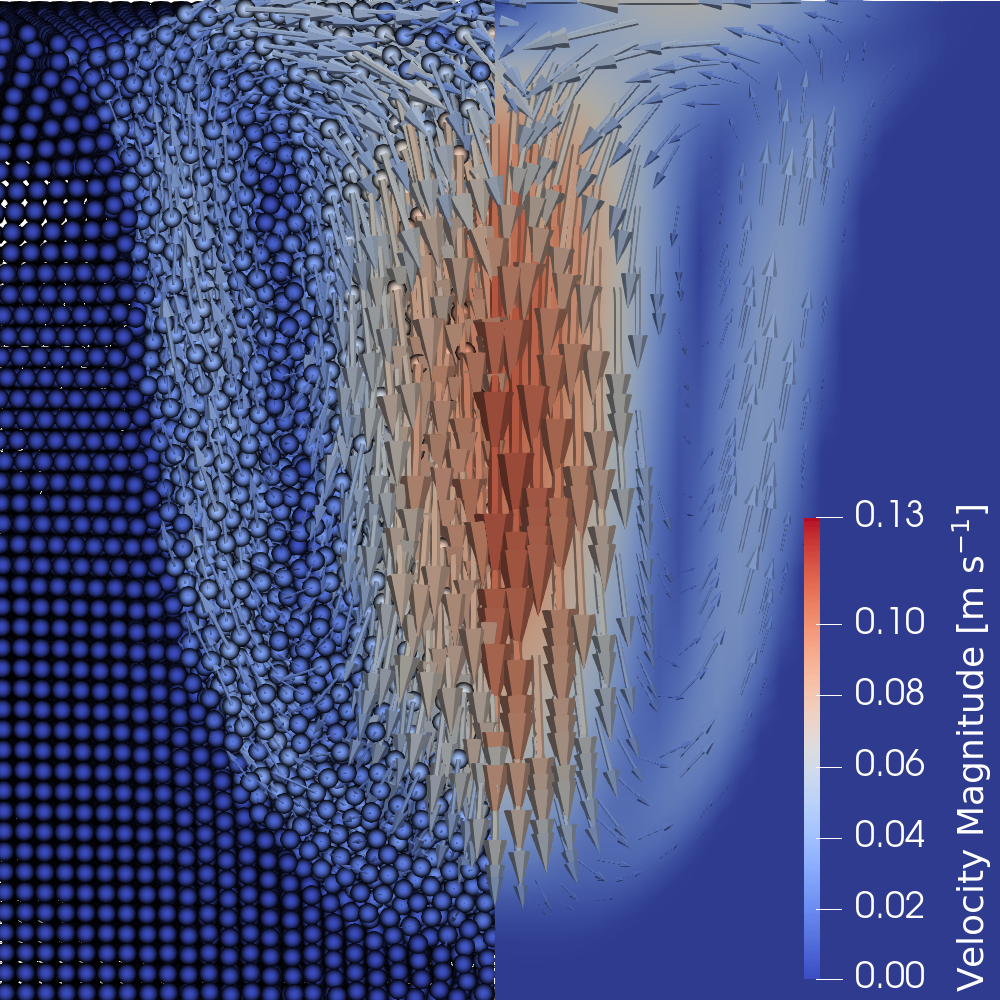}
    \caption{Velocity at t=1.6s.}
    \label{fig:crosssection_velocity_1.6}
\end{subfigure}
\caption{Side-by-by comparison of SPH and FEM results of temperature (left column) and velocity (right column).
The SPH solution is always shown on the left within each figure while the WM solution is shown on the right.}
\end{figure*}

Lastly, the temperature for the chosen characteristic time steps along the symmetry axis (z) is shown in Figure \ref{fig:temp_height_axis}.
Again similar observations regarding agreement as before can be made. 
We see excellent agreement in the temperature distribution for $t=1.0$\si{\s} between all three methods, while some deviation can be observed for times close to the beginning $t=0.3$\si{\s} and for times close to full penetration $t=1.6$\si{\s}.
At time $t=0.3$\si{\s} it should be pointed out that SPH reaches a higher maximal temperature at $z=0.005$\si{\m} and does not share the characteristic shape we observe for the other distributions, but exhibits a more uniform temperature decay as one would expect from a purely conductive heat transfer. 
As mentioned earlier in the description of Figure \ref{fig:temperature_contours}, this is most likely due to inhibited fluid movement at early times, as can also be observed later in Figure \ref{fig:crosssection_velocity_0.3} for $t = 0.3$\si{\s}.
Again, for this time step $t=0.3$s the COMSOL aligns more closely with the WM method, although not reaching perfect agreement. 
The shape of the COMSOL solution at this time seems to indicate weaker conductive heat transfer than WM, therefore most likely representing an intermediate state between the SPH and WM solution, which is an observation already made for some of the previous data, e.g. Figure \ref{fig:vel_height_axis}.

For $t=1.0$\si{\s} we see excellent agreement between all three solutions for the lower half of the temperature distributions ($z > 0.0025$\si{\m}), however in the upper half we observe slightly higher temperatures in SPH compared to WM and COMSOL.

At $t=1.6$\si{\s} the largest discrepancy between the three solutions can be found at the bottom of the melt-pool.  
While the SPH and COMSOL solution show very good agreement at the lower part of the axis ($z \leq 0.001$\si{\m}), the WM solution differs strongly from both. 
This could be, as previously mentioned, due to the effect of the lower boundary, as WM has already achieved full penetration at this time step, while the SPH and COMSOL solution have not yet achieved it. 
However, for the upper part of the axis ($z > 0.001$\si{\m}), the agreement between COMSOL and WM is very good, while the temperatures calculated by SPH is slightly higher, with the deviation further increasing slightly with increasing z. 
All distributions have the previously mentioned characteristic shape, however the WM solution consistently predicts a deeper weld pool, indicating a stronger convective heat transfer in the entire melt-pool, consistent with the observations on the velocities in Figure \ref{fig:vel_height_axis}.
Regarding the temperature profile along the height axis it can also be said that our SPH method lies well within the margin of error that can be expected between two different FEM solvers.

\paragraph{Qualitative Comparison}

Following the quantitative an\-alysis we will now show a qualitative comparison of the SPH and COMSOL solutions for the half of the cross section at times $t$ = \SI{0.3}{\second}, $t$ = \SI{1.0}{\second} and $t$ = \SI{1.6}{\second}.
These times were selected, as both $t$ = \SI{0.3}{\second} as well as $t$ = \SI{1.6}{\second} showed the greatest difference between the methods with regards to the melt-pool dimensions shown in Figure \ref{fig:pool_dimensions}. While the deviations were strongest for the comparison with the WM solution, the qualitative comparison will compare the SPH and the COMSOL solution where the agreement with of the melt-pool dimensions was best.

Over the whole duration, the time $t$ = \SI{0.3}{\second} showed the larges disagreement between both methods, as can be seen also from the study of Figure \ref{fig:pool_dimensions} and Figure \ref{fig:temperature_contours}. While the disagreement is within the expected margin of error given by the comparison of the two FEM methods, looking in more detail at Figure \ref{fig:crosssection_temperature_0.3} and Figure \ref{fig:crosssection_velocity_0.3}, one can see that the melt-pool (as indicated by the melting temperature iso-countour as a black line) in the COMSOL simulation is slightly deeper than the melt-pool of the SPH simulation, while the melt-pool radius appears comparable.
It is also possible to observe the distinct deviation in the velocity field between the two simulations, which is in line with the observation of the velocity in Figure \ref{fig:vel_height_axis}.
A stable vortex seems to have already developed in the COMSOL simulation, while the SPH particles are barely in motion.
We explain this by the very restrictive boundary condition in the SPH simulation.
Until there is enough space for particles to overwhelm, or circumvent, the pressure forces keeping the fluid incompressible, the fluid will remain motionless.
Since we are using an implicit pressure solver, the pressure forces will be quite large until a suitable large region of the fluid has melted, allowing for more movement while still maintaining incompressiblity.
The difference in velocity also explains the deviation in melt-pool shape, since at $t$ = \SI{0.3}{\second} the SPH simulation is still only transferring heat by conduction, while the melt-pool of the FEM simulation is increasing in depth due to an increase in convection.

Next, in Figure \ref{fig:crosssection_temperature_1.0} and Figure \ref{fig:crosssection_velocity_1.0}, we show the temperature and velocity distributions at $t$ = \SI{1.0}{\second} for both SPH and COMSOL.
Overall these results show excellent agreement, both in terms of the melt-pool dimensions (width and height) as well as the actual temperature and velocity field.
The temperature fields appear almost identical, however, the melt-pool contour exhibits small deviations, especially close to the top boundary.
In addition, the velocity magnitude of the SPH simulation at the symmetry axis is slightly lower than that of the COMSOL simulation.
We are unsure about this deviation and can only speculate that due to the increased pressure forces (slight over-pressure) as mentioned above, the velocity slightly decelerated.
Otherwise, there are no movement inhibiting forces acting on the fluid which are not shared across both simulations, i.e. viscosity and momentum-sink.
Regardless of the difference in velocity magnitude, the flow pattern is, similar to the temperature field, almost identical.

Finally, considering the simulations at $t$ = \SI{1.6}{\second} in Figure \ref{fig:crosssection_temperature_1.6} and Figure \ref{fig:crosssection_velocity_1.6} we show the instance where full penetration is reached for both solvers, almost simultaneously, see also Figure \ref{fig:pool_dimensions}. 
One can see that the comparison of the temperature fields show very good agreement as well, where the general shape of the melting temperature iso-countour is captured accurately almost in the entire region, except for a small area close to the top surface.
The velocity field also exhibits the same flow pattern as at $t$ = \SI{1}{\second}, where the velocity magnitude is again slightly lower in the SPH simulation than in the COMSOL simulation.

Overall, we can observe very good agreement between the flow patterns and temperature distributions of both simulation methods.
We speculate that the observed differences at $t$ = \SI{0.3}{\second} results from restricted motion in SPH due to high pressure forces, which would also explain the diminished velocity magnitude and subsequent time steps, but could not confirm nor deny this speculation in a brief investigation.
While this discrepancy warrants further study, we have shown successfully that the observed flow pattern and temperature field agree very well over a long duration and retain distinct similarities when the melt-pool depth differs.

\section{Discussion}
\label{sec:discussion}
The comparison of the results from the SPH, WM and COMSOL solvers yields intriguing results. 
The simplifications strongly favor the application of an Eulerian method, and one of the main goals of this comparison was to investigate whether the SPH method is able to achieve good results in such a setting.
Under consideration of the previous evaluation we believe this to absolutely be the case, as we were able to observe very good quantitative agreement between the SPH and Comsol methods, while observing the same overall trends with the WM method.

Firstly, the conductive heat transfer was confirmed to work as expected, up to differences which can be accounted for by differences in spatial and temporal discretization. 
Secondly, the convective heat transfer seems to be in good agreement between SPH and COMSOL as could be seen for example by the melting temperature iso-contours of Figure \ref{fig:contour_noEM}. The convective heat transfer in WM seemed to occur slightly faster.
This is in line with the observations on the velocities, see Figure \ref{fig:vel_height_axis}, that seem to exhibit some deviation for all three methods of $\approx \pm 20\%$. In the light of these deviations the excellent agreement for the melt pool depth at $0.6$\si{\s}$<t<1.3$\si{\s} seems somewhat surprising.
Figure \ref{fig:temp_height_axis} seems to confirm that the convective heat transfer in downward direction is slightly slower at all times of the SPH simulations as the temperature at the top, close to the heat source, is slightly higher.
Nevertheless, the temperature profiles align almost perfectly for SPH and COMSOL towards the bottom of the melt pool, where the WM solution deviates significantly from both.
The movement of the melt pool depth in SPH and COMSOL between $0.2$\si{\s}$<t<0.5$\si{\s}, as seen in Figure \ref{fig:pool_dimensions} seems especially intriguing, as both methods agree while the WM solution shows a deviation. 
We can only assume that this deviation between the two FEM solvers is due to implementation specific details.

Another observation relates to the fact that the SPH solution deviates from the other two mostly for earlier times, i.e. $t<0.3$\si{\s}, where the number of particles in the melt pool seems to be insufficient to allow for a development of a flow pattern, while at the same time the implementation of a fixed boundary wall is not native to SPH, as SPH has its strength especially in the treatment of unbounded domains, and to a certain degree free fluid surfaces. 
The pressure forces in SPH might actually block the flow, while for WM the consideration of fixed wall boundary conditions is natural to the method and no such restrictions exist when a free-slip condition is used. 

It should be kept in mind however, that for a real process, the free surface is not actually confined in such a way. 
Of course, for a comparison of melt pool development with the real process, also the non-linearity of the material parameters must be considered. 
Here, the inclusion of the latent heat in terms of the Stefan-Problem is known to present a significant challenge, among others.
Nevertheless, even established Eulerian approaches have difficulty reaching satisfactory agreement without the use of strong compensation models and all the peculiarities that derive from the challenges related to the free surface deformation, discontinuities and strong topological changes, that are typical for arc welding processes.
As such, while the inclusion of latent heat in terms of the Stefan-Problem is also no trivial task for SPH methods, we consider our SPH model to be a suitable candidate for performing simulations of the more complex interactions and physics of the real process, due to the improved ability to model surface deformation, discontinuities and significant topological changes.

\section{Conclusion}
\label{sec:conclusion}
In this work a full 3D SPH model for the simulation of a TIG spot-welding process was proposed and compared quantitatively with a 2+1D FEM simulation implemented in Wolfram Mathematica and COMSOL.
We were able to show a remarkable agreement of our SPH method and the WM and COMSOL methods, especially given the differences that we were able to observe between the WM-FEM and the COMSOL-FEM solving entirely identical problems.

Overall, our work confirms the expectation that the SPH method represents an at least equivalent capability of quantitatively modeling welding processes. 
This holds even under conditions that are simplified in such a way to compensate for the limitations of the Eulerian methods (free-surface), while at the same time strongly favoring the application of a Eulerian method (bounded domain). 
Therefore, we expect that the SPH method can outperform the approach posed by Eulerian methods under more realistic conditions, i.e. allowing for a free surface, and requiring full 3D also for the Eulerian Method, especially once the numerical treatment of the non-linear material parameters has been realized.

\section*{Acknowledgements}
The presented investigations were carried out at RWTH Aachen University within the framework of the Collaborative Research Centre SFB1120-236616214 “Bau\-teilpräzision durch Beherrschung von Schmelze und Erstarrung in Produktionsprozessen” and funded by the Deutsche Forschungsgemeinschaft e.V. (DFG, German Research Foundation). The sponsorship and support is gratefully acknowledged. The COMSOL and WM models were developed in project DFG-390246097, funded by the Deutsche Forschungsgemeinschaft e.V. (DFG, German Research Foundation). Simulations were performed in part with computing resources granted by RWTH Aachen University under project rwth0398.

%
\section*{Conflict of interest}

On behalf of all authors, the corresponding author states that there is no conflict of interest.

\bibliographystyle{spmpsci}      
\bibliography{references_cleaned.bib}   

\begin{thebibliography}{10}
\providecommand{\url}[1]{{#1}}
\providecommand{\urlprefix}{URL }
\expandafter\ifx\csname urlstyle\endcsname\relax
  \providecommand{\doi}[1]{DOI~\discretionary{}{}{}#1}\else
  \providecommand{\doi}{DOI~\discretionary{}{}{}\begingroup
  \urlstyle{rm}\Url}\fi

\bibitem{AIA+12}
Akinci, N., Ihmsen, M., Akinci, G., Solenthaler, B., Teschner, M.: {Versatile
  rigid-fluid coupling for incompressible SPH}.
\newblock ACM Transactions on Graphics \textbf{31}(4), 1--8 (2012).
\newblock \doi{10.1145/2185520.2335413}.
\newblock \urlprefix\url{http://dl.acm.org/citation.cfm?doid=2185520.2335413}

\bibitem{BTN13}
Barecasco, A., Terissa, H., Naa, C.F.: Simple free-surface detection in two and
  three-dimensional sph solver  (2013)

\bibitem{BT07}
Becker, M., Teschner, M.: {Weakly compressible SPH for free surface flows}.
\newblock In: ACM SIGGRAPH/Eurographics Symposium on Computer Animation, pp.
  1--8 (2007).
\newblock
  \urlprefix\url{http://portal.acm.org/citation.cfm?id=1272690.1272719{\%}5Cnhttp://dl.acm.org/citation.cfm?id=1272719}

\bibitem{splishsplash}
Bender, J.: {SPlisHSPlasH} (2021).
\newblock
  \urlprefix\url{{https://github.com/InteractiveComputerGraphics/SPlisHSPlasH}}

\bibitem{BK15}
Bender, J., Koschier, D.: {Divergence-Free Smoothed Particle Hydrodynamics}.
\newblock In: ACM SIGGRAPH/Eurographics Symposium on Computer Animation, pp.
  1--9 (2015)

\bibitem{BKWK19}
Bender, J., Kugelstadt, T., Weiler, M., Koschier, D.: Volume maps: An implicit
  boundary representation for sph.
\newblock In: Motion, Interaction and Games, p.~26. ACM (2019)

\bibitem{BVR88}
Brent, A.D., Voller, V.R., Reid, K.J.: Enthalpy-porosity technique for modeling
  convection-diffusion phase change: Application to the melting of a pure
  metal.
\newblock Numerical Heat Transfer \textbf{13}(3), 297--318 (1988).
\newblock \doi{10.1080/10407788808913615}

\bibitem{Bro85}
Brookshaw, L.: A method of calculating radiative heat diffusion in particle
  simulations.
\newblock Publications of the Astronomical Society of Australia \textbf{6}(2),
  207--210 (1985).
\newblock \doi{10.1017/s1323358000018117}

\bibitem{CCC+20}
Cadiou, S., Courtois, M., Carin, M., Berckmans, W., {Le masson}, P.: 3d heat
  transfer, fluid flow and electromagnetic model for cold metal transfer wire
  arc additive manufacturing (cmt-waam).
\newblock Additive Manufacturing \textbf{36}, 101541 (2020).
\newblock \doi{10.1016/j.addma.2020.101541}.
\newblock
  \urlprefix\url{https://www.sciencedirect.com/science/article/pii/S2214860420309131}

\bibitem{CN21}
Cho, W.I., Na, S.J.: Impact of driving forces on molten pool in gas metal arc
  welding.
\newblock Welding in the World  (2021).
\newblock \doi{10.1007/s40194-021-01138-8}

\bibitem{DC16}
Das, R., Cleary, P.: Three-dimensional modelling of coupled flow dynamics, heat
  transfer and residual stress generation in arc welding processes using the
  mesh-free {SPH} method.
\newblock Journal of Computational Science \textbf{16}, 200--216 (2016).
\newblock \doi{10.1016/j.jocs.2016.03.006}

\bibitem{davidson2001}
Davidson, P.: An Introduction to Magnetohydrodynamics.
\newblock Cambridge University Press (2001)

\bibitem{GM77}
Gingold, R.a., Monaghan, J.: {Smoothed Particle Hydrodynamics: Theory and
  Application to Non-Spherical Stars}.
\newblock Monthly Notices of the Royal Astronomical Society (181), 375--389
  (1977).
\newblock \doi{10.1093/mnras/181.3.375}

\bibitem{haidar1996}
Haidar, J., Lowke, J.: Predictions of metal droplet formation in arc welding.
\newblock Journal of Physics D: Applied Physics \textbf{29}(12), 2951 (1996)

\bibitem{hertel2014}
Hertel, M., F{\"u}ssel, U., Schnick, M.: Numerical simulation of the
  plasma--mig process—interactions of the arcs, droplet detachment and weld
  pool formation.
\newblock Welding in the World \textbf{58}(1), 85--92 (2014)

\bibitem{IIFS11}
Ito, M., Izawa, S., Fukunishi, Y., Shigeta, M.: Sph simulation of gas arc
  welding process.
\newblock In: Proc. Seventh International Conference on Computational Fluid
  Dynamics (Hawaii, 2012), ICCFD7-3706 (2012).
\newblock
  \urlprefix\url{{https://iccfd.org/iccfd7/assets/pdf/papers/ICCFD7-3706_paper.pdf}}

\bibitem{INI15}
Ito, M., Nishio, Y., Izawa, S., Fukunishi, Y., Shigeta, M.: Numerical
  simulation of joining process in a {TIG} welding system using incompressible
  {SPH} method.
\newblock {QUARTERLY} {JOURNAL} {OF} {THE} {JAPAN} {WELDING} {SOCIETY}
  \textbf{33}(2), 34s--38s (2015).
\newblock \doi{10.2207/qjjws.33.34s}.
\newblock
  \urlprefix\url{{https://www.jstage.jst.go.jp/article/qjjws/33/2/33_34s/_pdf/-char/en}}

\bibitem{KST18}
Komen, H., Shigeta, M., Tanaka, M.: Numerical simulation of molten metal
  droplet transfer and weld pool convection during gas metal arc welding using
  incompressible smoothed particle hydrodynamics method.
\newblock International Journal of Heat and Mass Transfer \textbf{121},
  978--985 (2018).
\newblock \doi{10.1016/j.ijheatmasstransfer.2018.01.059}

\bibitem{KTT20}
Komen, H., Tanaka, M., Terasaki, H.: Three-dimensional simulation of gas metal
  arc welding process using particle-grid hybrid method.
\newblock {QUARTERLY} {JOURNAL} {OF} {THE} {JAPAN} {WELDING} {SOCIETY}
  \textbf{38}(2), 25s--29s (2020).
\newblock \doi{10.2207/qjjws.38.25s}.
\newblock
  \urlprefix\url{{https://www.jstage.jst.go.jp/article/qjjws/38/2/38_25s/_pdf/-char/en}}

\bibitem{KBST19}
Koschier, D., Bender, J., Solenthaler, B., Teschner, M.: Smoothed particle
  hydrodynamics techniques for the physics based simulation of fluids and
  solids.
\newblock In: EUROGRAPHICS 2019 Tutorials. Eurographics Association (2019)

\bibitem{KS85}
Kou, S., Sun, D.K.: Fluid flow and weld penetration in stationary arc welds.
\newblock Metallurgical Transactions A \textbf{16}(2), 203--213 (1985).
\newblock \doi{10.1007/bf02816047}

\bibitem{Luc77}
Lucy, L.B.: A numerical approach to the testing of the fission hypothesis.
\newblock Astronomical Journal \textbf{82}, 1013--1024 (1977).
\newblock \doi{10.1086/112164}

\bibitem{medale2004}
Medale, M., Rabier, S., Xhaard, C.: A thermo-hydraulic numerical model for high
  energy welding processes.
\newblock Revue Europ{\'e}enne des El{\'e}ments \textbf{13}(3-4), 207--229
  (2004)

\bibitem{medale2008}
Medale, M., Touvrey, C., Fabbro, R.: An axi-symmetric thermo-hydraulic model to
  better understand spot laser welding.
\newblock European Journal of Computational Mechanics/Revue Europ{\'e}enne de
  M{\'e}canique Num{\'e}rique \textbf{17}(5-7), 795--806 (2008)

\bibitem{MSS+21}
Mokrov, O., Simon, M., Shvartc, I., Sharma, R., Reisgen, U.: Validation of the
  {EDACC} model for {GMAW} process simulation by weld pool dimension
  comparison.
\newblock In: Lecture Notes in Mechanical Engineering, pp. 51--59. Springer
  International Publishing (2021).
\newblock \doi{10.1007/978-3-030-70332-5_5}

\bibitem{nguyen2017}
Nguyen, M.C., Medale, M., Asserin, O., Gounand, S., Gilles, P.: Sensitivity to
  welding positions and parameters in gta welding with a 3d multiphysics
  numerical model.
\newblock Numerical Heat Transfer, Part A: Applications \textbf{71}(3),
  233--249 (2017)

\bibitem{Pri12a}
Price, D.J.: Smoothed particle hydrodynamics and magnetohydrodynamics.
\newblock Journal of Computational Physics \textbf{231}(3), 759--794 (2012).
\newblock \doi{10.1016/j.jcp.2010.12.011}

\bibitem{schenk2004}
Schenk, O., G{\"a}rtner, K.: Solving unsymmetric sparse systems of linear
  equations with pardiso.
\newblock Future Generation Computer Systems \textbf{20}(3), 475--487 (2004)

\bibitem{SDK+12}
Semenov, O., Demchenko, V., Krivtsun, I., et~al.: Modelling of the droplet
  formation process in {GMA} welding.
\newblock Proc. of 10th Int. Sem. on Numerical Analysis of Weldability pp.
  83--94 (2012)

\bibitem{semenov2021}
Semenov, O., Krivtsun, I., Lykhoshva, A., Hluchenkyi, O., Bondar, O.:
  Comparative analysis of the results of computer simulation of heat transfer
  and hydrodynamic processes in the metal being welded by means of different
  software tools.
\newblock Paton Welding Journal (01), 20--24 (2021).
\newblock \doi{10.37434/as2021.01.04}

\bibitem{semenov2014}
Semyonov, A.: Methods of mathematical modelling of the processes of electrode
  metal drop formation and transfer in consumable electrode welding.
\newblock Paton Welding Journal (10), 2--10 (2014)

\bibitem{THF17}
Trautmann, M., Hertel, M., Füssel, U.: Numerical simulation of {TIG} weld pool
  dynamics using smoothed particle hydrodynamics.
\newblock International Journal of Heat and Mass Transfer \textbf{115},
  842--853 (2017).
\newblock \doi{10.1016/j.ijheatmasstransfer.2017.08.060}

\bibitem{THF18}
Trautmann, M., Hertel, M., Füssel, U.: Numerical simulation of weld pool
  dynamics using a {SPH} approach.
\newblock Welding in the World \textbf{62}(5), 1013--1020 (2018).
\newblock \doi{10.1007/s40194-018-0615-5}

\bibitem{VBP90}
Voller, V., Brent, A., Prakash, C.: Modelling the mushy region in a binary
  alloy.
\newblock Applied Mathematical Modelling \textbf{14}(6), 320--326 (1990).
\newblock \doi{10.1016/0307-904x(90)90084-i}

\bibitem{WKBB18}
Weiler, M., Koschier, D., Brand, M., Bender, J.: A physically consistent
  implicit viscosity solver for sph fluids.
\newblock CGF \textbf{37}(2) (2018)

\bibitem{zienkiewicz2014}
Zienkiewicz, O., Taylor, R., Nithiarasu, P.: The Finite Element Method for
  Fluid Dynamics.
\newblock Butterworth-Heinemann, London (2014)

\bibitem{zienkiewicz1995}
Zienkiewicz, O.C., Codina, R.: A general algorithm for compressible and
  incompressible flow—part i. the split, characteristic-based scheme.
\newblock International Journal for Numerical Methods in Fluids
  \textbf{20}(8-9), 869--885 (1995)

\end{thebibliography}

\end{document}